\documentclass[twocolumn, tighten, numberedappendix]{aastex631}
\pdfoutput=1 

\usepackage{amsmath,amstext}
\usepackage[T1]{fontenc}
\usepackage{apjfonts} 
\usepackage{graphicx}
\usepackage{xcolor}
\usepackage{color}
\usepackage{appendix}
\usepackage{microtype}
\usepackage{comment}

\usepackage{savesym}
\savesymbol{tablenum}
\usepackage[separate-uncertainty=true, multi-part-units=single]{siunitx}
\restoresymbol{SIX}{tablenum}

\usepackage{gensymb}
\usepackage{float}
\usepackage{makecell}
\usepackage{hyperref}
\usepackage[hyphens]{url}

\providecommand{\sorthelp}[1]{}

\usepackage{natbib}

\shorttitle{}

\begin{document}

\title{Limits on Polarized Dust Spectral Index Variations for CMB Foreground Analysis}

\correspondingauthor{Keisuke Osumi}
\email{kosumi1.jhu.edu}

\author[0000-0003-2838-1880]{Keisuke Osumi}
\affiliation{Johns Hopkins University \\
3400 North Charles Street \\
Baltimore, MD 21218, USA}

\author[0000-0003-3017-3474]{Janet L. Weiland}  
\affiliation{Johns Hopkins University \\
3400 North Charles Street \\
Baltimore, MD 21218, USA}

\author[0000-0002-2147-2248]{Graeme E. Addison}
\affiliation{Johns Hopkins University \\
3400 North Charles Street \\
Baltimore, MD 21218, USA}

\author[0000-0001-8839-7206]{Charles L. Bennett}
\affiliation{Johns Hopkins University \\
3400 North Charles Street \\
Baltimore, MD 21218, USA}

\begin{abstract}
Using \textit{Planck} polarization data, we search for and constrain spatial variations of the polarized dust foreground for cosmic microwave background (CMB) observations, specifically in its spectral index, $\beta_d$.
Failure to account for such variations will cause errors in the foreground cleaning that propagate into errors on cosmological parameter recovery from the cleaned CMB map.
It is unclear how robust prior studies of the \textit{Planck} data which constrained $\beta_d$ variations are due to challenges with noise modeling, residual systematics, and priors.
To clarify constraints on $\beta_d$ and its variation, we employ two pixel space analyses of the polarized dust foreground at $>3.7^{\circ}$ scales on $\approx 60\%$ of the sky at high Galactic latitudes.
A template fitting method, which measures $\beta_d$ over three regions of $\approx 20\%$ of the sky, does not find significant deviations from an uniform $\beta_d = 1.55$, consistent with prior \textit{Planck} determinations.
An additional analysis in these regions, based on multifrequency fits to a dust and CMB model per pixel, puts limits on $\sigma_{\beta_d}$, the Gaussian spatial variation in $\beta_d$.
At the highest latitudes, the data support $\sigma_{\beta_d}$ up to $0.45$, $0.30$ at mid-latitudes, and $0.15$ at low-latitudes.
We also demonstrate that care must be taken when interpreting the current \textit{Planck} constraints, $\beta_d$ maps, and noise simulations.
Due to residual systematics and low dust signal to noise at high latitudes, forecasts for ongoing and future missions should include the possibility of large values of $\sigma_{\beta_d}$ as estimated in this paper, based on current polarization data.
\end{abstract}


\section{Introduction}
\label{sec:intro}
Since its discovery by \cite{PW1965}, the cosmic microwave background (CMB) has played a central role in cosmology.
The CMB intensity and its anisotropy have been well studied and have provided some of the strongest evidence and constraints for the prevailing $\Lambda$CDM model of the universe \citep[e.g.][]{WMAP2013, P2018_params}.
Measuring the CMB polarization improves these constraints.
In particular, the CMB polarization anisotropy can be used to constrain reionization \citep[e.g.][]{Page2007, P2018_params, Pagano2020, Watts2020} and provide an indirect detection of primordial gravitational waves \citep[e.g.][]{kamionkowski1997}.
For this reason, there are a number of existing and future experiments designed to undertake increasingly challenging measurements of the CMB polarization anisotropy\footnote{A list of experiments and their properties is provided at: \href{https://lambda.gsfc.nasa.gov/product/expt/}{https://lambda.gsfc.nasa.gov/product/expt/}}.

One of the major analysis challenges for CMB experiments is that of separating foreground and CMB components.  
This is especially true for polarized observations, where emission from Galactic synchrotron and thermal dust emission dominates the fainter polarized CMB signal, even at high Galactic latitudes.  
We focus on the polarized dust foreground, for which the \textit{Planck} multifrequency high frequency instrument (HFI) data provide the current best high latitude sky coverage.  

The observed dust signal is a superposition of emission from the dust grain population along the line of sight, which includes grains with differing sizes, shapes, compositions, subject to ambient radiation fields, all of which affect the steady-state dust temperature, $T_d$, and the observed spectral energy distribution (SED) \citep[e.g.][]{Draine1984, Weingartner2001}.
Aspherical dust grains tend to be aligned with the longest axis perpendicular to the local magnetic field and their most efficient emission or absorption is along the longest axis, which produces an overall polarization \citep[e.g.][]{Lazarian2003, Lazarian2009}.
This alignment, and hence the polarization, also depends on the properties of the dust grains \citep{Hildebrand2000}.

Given the complexity of physics leading to the observed dust signal along a line of sight, one could attempt to fit a multi-component model with a range of dust populations with different dust temperatures, $T_d$, and spectral indices, $\beta_d$.
A priori, it would not be surprising if $\beta_d$ varies significantly across the sky as different dust populations with different compositions of dust grains are considered.  
However, \textit{Planck} measurements indicate a large degree of uniformity of the dust foreground across most of the sky.  
The \textit{Planck} Commander pixel-based analysis for public data release 3 (PR3) found a value of $\beta_d = 1.6 \pm 0.13$ over the sky \citep{P2018_foregrounds}.
Similar results were found for the PR4 foreground analyses \citep{P2020}.
Based on these findings, \citet{P2018_foregrounds} and \citet{P2020} adopted a Gaussian prior of $\beta_d = 1.6 \pm 0.1$ for the fit in their final foreground model.
However, when a Gaussian prior is omitted from the fit, large tails are present in the recovered $\beta_d$ distributions in Figure 29 of \citet{P2018_foregrounds}. These were ascribed to poor fits in noise dominated regions of the sky, but could also be due to spatial variation in $\beta_d$. 

A more focused study of the dust foreground was conducted on the older \textit{Planck} PR2 dataset in \citet{P2015_dust}. 
This study considered $\beta_d$ on disjoint $10^\circ$ circular patches, but treated each disjoint patch as an independent measurement of a spatially uniform $\beta_d$ and did not attempt to characterize spatial variations in $\beta_d$. 

\citet{P2018_dust} also conducted a focused study of the dust foreground for the \textit{Planck} PR3 dataset, which significantly reduced the level of residual systematics. 
They computed the angular power spectra in six overlapping nested regions with effective sky fractions of 24 to 71\% to determine a constraint in the allowed Gaussian variation in $\beta_d$ of $\sigma_{\beta_d} = 0.092 \pm 0.005$. 
However, this constraint is defined as the standard deviation over the six highly correlated $\beta_d$ estimates from their six nested regions, which limits its applicability.

Spatial variations in $\beta_d$ can also produce a decorrelation of the BB power spectrum between the \textit{Planck} 217 GHz and 353 GHz polarization data, which is expected to be more significant at smaller ($\ell > 50$ or $<3.7^{\circ}$) scales than we investigate in this work.
This was explored in \citet{P2016_dust}, \citet{Sheehy2018}, and \citet{P2018_dust}.
\citet{Sheehy2018} and the later \textit{Planck} analysis in \citet{P2018_dust} do not find a significant decorrelation in the B-modes.
Upon including decorrelation in their CMB analysis, \citet{BICEP2018} found that their constraints are not significantly shifted and find little evidence for decorrelation, at least given the current experimental noise.

In the analyses cited above, it is possible to achieve stronger constraints on $\beta_d$ and its spatial variation by introducing intensity data, since the dust intensity has a higher signal to noise ratio.
Although \cite{P2018_dust} found that the mean sky $\beta_d$ was consistent between polarization and intensity, differences can arise for example by magnetic dipole emission due to inclusions randomly-oriented within interstellar dust grains, which can reduce the polarization of the dust emission \citep{2013_Draine}.

Several studies have shown evidence for spatial variation in $\beta_d$.
Recently, \citet{Pelgrims2021} found that combining \textit{Planck} polarization data with neutral hydrogen measurements and leveraging a frequency decorrelation analysis can show detectable differences in the dust SED along specific lines of sight targeted by HI observations. 
As mentioned earlier, other analyses conducted over large sky regions also show hints of spatial variation in $\beta_d$, but include complicating factors beyond instrumental noise like the choice of noise model, prior choices, and residual systematics \citep[e.g.][]{P2015_dust, P2016_dust, P2015_foregrounds, P2018_foregrounds, P2018_dust, P2020}.
These additional features are difficult to disentangle from true on-sky $\beta_d$ variations.

To shed some light on this question, we engage in a pixel space investigation of the power the \textit{Planck} polarization data have to constrain $\beta_d$ variations on $\approx 60\%$ of the sky at high Galactic latitudes.
We chose to work in pixel space since both the dust foreground signal and the primary challenges we face in this analysis (residual systematics and the accuracy of the noise model), have large scale spatial structures that are easier to account for in pixel space.
Power spectra of foreground signals lose information and the data compression it provides is not necessary in this analysis, as we restrict ourselves to large angular scales at low resolution ($>3.7^{\circ}$).

This paper is structured as follows.
Section~\ref{sec:data} specifies the \textit{Planck} data products used.
Section~\ref{sec:model} describes the foreground model and how the realistic simulations were constructed.
Section~\ref{sec:TFmethod} presents the first analysis method, which is based on a two band template fit, with corresponding results included in Section~\ref{sec:TFresults}.
Section~\ref{sec:fitting_method} describes the more flexible second analysis we conducted, which was based on a multifrequency fit of the model to the data, with a description of results in Section~\ref{sec:fitting_results}.
These include a comparison between the simulations and the data, with findings on the power of the \textit{Planck} polarization data to constrain $\beta_d$ variations. 
Finally, we conclude in Section~\ref{sec:conc}.

We provide a roadmap to this paper in Table~\ref{tab:roadmap}, which may be a useful guide to the reader.

\begin{table*}[t]
\centering
\begin{tabular}{c  c  c  c  c} 
 \hline 
 Sec. & Fig./Table & Analysis & $\beta_d$ & Bands \\ 
 \hline\hline
  & & Two band template fit & & \\
 \hline
 \ref{sec:TFresults} & Table~\ref{tab:TFres} & Dust template fit after CMB removal & Solved & 217 GHz \& 353 GHz\\
                     &                       & over 3 disjoint regions &        & \\
 \hline\hline
  & & Multifrequency fit & & \\
  \hline
  \ref{ssec:dvs_comps} & Fig.~\ref{fig:dvs_res} & KS tests for $\chi^2$ PTE consistency & 1.55 fixed & 100-353 GHz \\
                       & Table~\ref{tab:KStest_dvs}~\&~\ref{tab:PTEsummary}  & between data and simulations      & & \\
  \hline
  \ref{ssec:dvs_comps} & Fig.~\ref{fig:dvs_cor}     & Residual-353 GHz dust correlation for & 1.55 fixed & 100-353 GHz \\
                       & Table~\ref{tab:PTEsummary} &  PTE consistency between data and sims &           & \\
  \hline
  \ref{ssec:bd_offset} & Fig.~\ref{fig:bdstep_cor}  & Residual-353 GHz dust correlation & 1.35 - 1.75 stepped & 100-353 GHz \\
                       &                            & varying fixed $\beta_d$ on full sky &           & \\
  \hline
  \ref{ssec:bdvary}    & Fig.~\ref{fig:KStest_bdvary}& Injecting random Gaussian variation & Gausian mean = 1.55 & 100-353 GHz \\
  \hline
  \ref{ssec:bd_inject} & Fig.~\ref{fig:com_gnilc_bd_maps}& Injecting GNILC or Commander variations & Fixed $\beta_d$ and $T_d$ from & 100-353 GHz \\
                       & Table~\ref{tab:PTEsummary} &                                   & GNILC or Commander & \\
\end{tabular}
\caption{Roadmap of analyses in this paper}
\label{tab:roadmap}
\end{table*}

\section{Data and masks}
\label{sec:data}
\subsection{Data}
\label{ssec:data}
In this work, we analyze the 2018 PR3 \textit{Planck} Stokes Q and U maps at the 4 highest polarized frequency channels from the \textit{Planck} High Frequency Instrument (HFI) at 100~GHz, 143~GHz, 217~GHz, and 353~GHz, using applicable correction maps, and their respective noise products, which include variance maps and noise simulations \citep{P2018_HFI}.
We use the \textit{Planck} 30~GHz polarization Q and U maps as a template to remove the synchrotron emission present in the HFI frequencies, and exclude the Low Frequency Instrument (LFI) 44 and 70~GHz channels from the analysis.
As discussed in Section~\ref{ssec:dvs_comps}, LFI data provide little constraining power on the synchrotron spectral index at the pixel scales used here.  
Use of the template removal approach for synchrotron therefore has two advantages. 
First, it avoids effects of the poorly understood noise properties at 44~GHz \citep[e.g.][]{BPlanck_2020}.  
Second, it mitigates the potential introduction of spurious large scale power into the analyses from calibration systematics in the LFI bands  \citep{P2018_LFI, Weiland2018}.

Since the release of the PR3 products, there have been two new data releases using alternative mapmaking pipelines, SRoll2 \citep{2019_sroll2} and NPIPE \citep{P2020}.
The analyses in this paper rely in part on estimates of the contribution of the CMB to the dust component separation, which are dominated by data in the HFI frequency bands. 
The \textit{Planck} Collaboration provides a number of CMB component estimates based on the PR3 maps (see Section~\ref{ssec:addprod}).  
Equivalent products do not accompany the SRoll2 release, and there is a substantial CMB mapping transfer function associated with the NPIPE maps.  
Although there are also processing improvements associated with the SRoll2 and NPIPE products, we chose to proceed with PR3.

As common in studies of CMB foregrounds in the pixel domain, the analysis requires the maps at each frequency considered to be at a common resolution, as defined by the smoothing scale and the \texttt{HEALPix} \citep{2005_Healpix} $N_\mathrm{side}$ pixel resolution parameter.
Therefore, before conducting the foreground analysis, we smooth each \textit{Planck} data product to a common resolution of 2 degrees full-width at half maximum (FWHM) and downgrade as necessary to $N_\mathrm{side} = 16$.

The foreground analysis also requires a noise model per pixel per frequency, for which the default variance maps packaged with the Q and U maps are insufficient.
Although the \textit{Planck} processing pipeline removes most instrumental effects from the data, there are residual systematics that are not removed nor accounted for in the default variance maps.
These residual systematics are modeled in \textit{Planck}'s Full Focal Plane 10 (FFP10) noise simulations \citep{P2018_HFI}.
These are end-to-end simulations that include all significant systematic effects at the time-ordered-data level, e.g. the analog to digital converter non-linearity and the bolometer time response.
However, due to computational constraints, the FFP10 simulations omit several components like the effect of discrete point sources, glitching/deglitching, and far sidelobes.
There are 300 such simulations available at each frequency.

We leveraged these 300 simulations per frequency to estimate variance maps that accounted for known residual systematic effects.
This was done by replicating the treatment of the Q and U data maps on the FFP10 simulated noise Q and U maps to produce 300 FFP10 noise simulations at the analysis resolution.
By computing variances per pixel over these 300 downgraded FFP10 simulations, we estimated the QQ, UU, and QU variance maps required for the analysis.

These estimated variance maps alone do not capture deviations from zero-mean Gaussian noise in the data.
To account for such deviations, rather than generating new Gaussian noise maps we use the FFP10 noise simulations as the noise component in the suite of simulations we calibrate the analysis against. 
This procedure is described in Section~\ref{ssec:sims}.

\subsection{Additional \textit{Planck} products}
\label{ssec:addprod}
Portions of the analysis also make use of CMB and dust foreground products provided in \textit{Planck} public releases.
In particular, we utilize products from four different foreground analyses conducted by the $\textit{Planck}$ Collaboration: Commander, GNILC, SEVEM, and SMICA.
We use the CMB estimates from all four analyses and, since only Commannder and GNILC produce a dust model, the dust model from these two analyses.

The Commander approach fits a parametric model to the data using a Gibbs sampling Monte Carlo method \citep{2004_COM, 2017_COM}. 
This model includes terms for the CMB and foregrounds and can be performed in the map or angular power spectrum domain.
We use the Commander CMB and dust products packaged with the earlier PR2 data release \citep{P2016}, as the publicly available PR3 Commander products do not include maps of the dust temperature, which is necessary to compare the foreground results against.

The Generalized Needlet Internal Linear Combination (GNILC) approach extracts maps for each component in the data by computing the minimum-variance linear combination of input maps in a needlet spherical wavelet basis \citep{2011_GNILC}. 
GNILC extends the classic internal linear combination method \citep{wmap2003}, which can only extract a component that is spatially uniform and rigidly scales with frequency, to account for more complex spatial and spectral behavior which allows for fits for foreground models.
As the PR3 GNILC products do not include maps of the $T_d$ or $\beta_d$, we revert to the earlier PR2 GNILC CMB and dust products for these quantities.

The Spectral Estimation Via Expectation Maximization (SEVEM) approach uses templates to separate the CMB from the foregrounds \citep{sevem2003, sevem2008, 2012_SEVEM}. 
Foreground templates are generated using the highest and lowest frequency bands and are used to remove the foreground from the CMB dominated intermediate bands.
As a consequence, SEVEM does not output a dust model, but we make use of the PR3 SEVEM CMB maps.

SMICA or Spectral Matching Independent Component Analysis \citep{2008_SMICA} is a non-parameteric method that fits the amplitude and spectral parameters of the CMB and foregrounds in harmonic space.
These values are used to derive harmonic space weights that are used to combine the data maps to produce a CMB map.
As with SEVEM, we use the PR3 SMICA CMB maps in part of the analysis.

\subsection{Masks}
\label{ssec:masks}

\begin{figure}[t]
\centering
\includegraphics[width=1.0\linewidth]{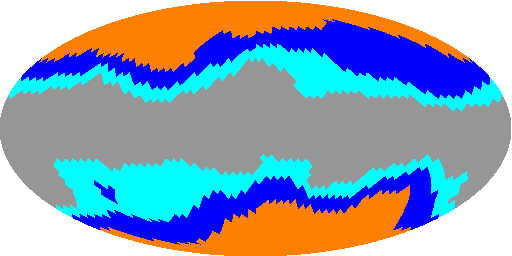}
\caption{We choose masks to define three disjoint sky regions that we analyze independently. Shown in Galactic coordinates, the high latitude mask (H) is orange, the mid-latitude mask (M) is blue, and the low-latitude mask (L) is cyan. The gray region is excluded from analysis. Each mask contains about 20\% of the sky for a total analyzed sky fraction of $\sim 0.6$. 
These masks sample thermal dust foreground regions with signal-to-noise ratio from lowest (orange) to highest (cyan), while avoiding complex foreground or residual structures closer to the Galactic plane (gray).
}
\label{fig:masks}
\end{figure}
We define three disjoint sky regions that we treat independently in the analyses.
These regions are chosen to probe different dust signal-to-noise ratio regimes, while excluding regions with complicated foreground structure and residual systematics along the Galactic plane. 
Each of these three regions covers approximately 20\% of the sky for a total analyzed sky fraction of about 0.6.

Figure~\ref{fig:masks} depicts the masks used to define the disjoint regions we consider in this work.
These masks are based on downgraded versions of those provided by \textit{Planck} and used in \cite{P2015_foregrounds}.
The original \textit{Planck} masks were produced by thresholding on the estimated total foreground amplitude in a pixel to target certain sky fractions.
The masks are produced by taking the difference between these \textit{Planck} masks from consecutive sky fractions.

\section{Model and simulations}
\label{sec:model}
\subsection{Model}
The Q and U data in each pixel for each frequency, $m(\nu)_{[Q,U]}$, can be modeled as a sum of the CMB, noise, and the dominant foregrounds in polarization: synchrotron and thermal dust.
This model, similar to that adopted by \textit{Planck} \citep{P2018_foregrounds}, represents the synchrotron as a power law in frequency, the dust as a modified blackbody, and holds the CMB component fixed across frequency, as the CMB is constant in the thermodynamic temperature units of the data:
\begin{align}
    m(\nu)_{[Q,U]} = & m_{\mathrm{synch}}(\nu)_{[Q,U]} + m_{\mathrm{dust}}(\nu)_{[Q,U]} + \text{CMB}_{[Q,U]}\nonumber\\
    = & A_{s, [Q, U]}\frac{g_s(\nu)}{g_s(\nu_s)}\left(\frac{\nu}{\nu_{s}}\right)^{\beta_s} \nonumber\\
    & + A_{d, [Q, U]}\frac{g_d(\nu)}{g_d(\nu_d)}\left(\frac{\nu}{\nu_{d}}\right)^{\beta_d+1}\frac{\exp({\frac{h\nu_d}{k_BT_d}}) -1}{\exp({\frac{h\nu}{k_BT_d}}) -1} \nonumber\\
    & + \text{CMB}_{[Q,U]}.\label{eqn:model}
\end{align}
Here, the subscript $[Q,U]$ denotes a 2 component vector for Q and U, $\nu$ is the frequency, $A_{s}$ is the synchrotron foreground amplitude, $A_d$ is the dust foreground amplitude, $\nu_{s}$ is the synchrotron reference frequency (30~GHz), $\nu_{d}$ is the dust reference frequency (353~GHz), $\beta_{s}$ is the synchrotron spectral index, $\beta_{d}$ is the dust spectral index, $T_d$ is the dust temperature, $h$ is Planck's constant, and $k_B$ is the Boltzmann constant.
The factors of $g_{fg}(\nu)$, where $f\!g$ refers to synchrotron ($s$) or dust ($d$), convert the brightness temperature units of the foreground models to the thermodynamic temperature units of the data.

Bandpass integrated color corrections, $C_{fg}(\nu)$, are needed to refer the true spectrum to the
adopted reference spectral dependence of the data. 
For synchrotron, we interpolate for $\beta_s = -3.1$ from a grid of correction values provided in the \textit{Planck} Explanatory Supplement\footnote{\url{https://wiki.cosmos.esa.int/planck-legacy-archive/index.php/UC_CC_Tables}}.
For dust, we use a \texttt{UC\_CC} IDL code derived in \cite{P2013HFIcalib} and provided by the \textit{Planck} collaboration\footnote{\url{https://wiki.cosmos.esa.int/planckpla2015/index.php/Unit\_conversion\_and\_Color\_correction}} to calculate $C_{d}(\nu)$ for a grid of varying $\beta_d$ and $T_d$ values which are interpolated over for each $\beta_d$ and $T_d$ value used in the fits. 
Additionally, conversion factors $U(\nu)$ are needed to convert from the brightness temperature units of the foreground model to  the thermodynamic temperature units of the data.
We use the values for $U(\nu)$ provided for each frequency in Table 2 of \cite{P2018_dust}, which were also derived using the same \texttt{UC\_CC} IDL code. 

Given $C_{fg}(\nu)$ and $U(\nu)$, the $g_{fg}(\nu)$ factors in Equation~\ref{eqn:model} above are defined as:
\begin{equation}
    g_{fg}(\nu) = \frac{C_{fg}(\nu)}{U(\nu)}.
\end{equation}

\subsection{Simulations}
\label{ssec:sims}
These analyses must be calibrated against accurate simulations of the \textit{Planck} data, which are known to contain non-Gaussianities and have a non-zero noise mean \citep{P2018_HFI}.

These simulations are generated at $N_{side} = 16$ and are defined as a sum of foregrounds, CMB, and noise.
This is the model as described in Equation~\ref{eqn:model}, but with an additional noise term.

The foreground component was constructed by taking the \textit{Planck} 30~GHz map as the synchrotron amplitude and the 353~GHz map as the dust amplitude and scaling to all intermediate frequencies assuming the \textit{Planck} best fit sky averages for $\beta_s = -3.1$, $\beta_d = 1.55$, and $T_d = 19.6\mathrm{K}$ \citep{P2018_foregrounds, P2018_dust}. 

The CMB component was generated with a fiducial input cosmology, with $10^9 A_s e^{-2\tau} = 1.881$, $\tau = 0.06$, and all other cosmological parameters set to those of the 2015F(CHM) (Plik) cosmology in table 6 of \cite{P2015_results}.
The noise component was set to be one of the downgraded FFP10 \textit{Planck} noise simulations rather than a Gaussian noise realization based on the noise variance estimates.
This allows us to take advantage of the \textit{Planck} collaboration's systematics modeling.
As there are only 300 FFP10 noise simulations available, we can produce only 300 baseline simulations.

Here, we have described the baseline simulations, but we also extend these simulations in Sections~\ref{ssec:bdvary} and~\ref{ssec:bd_inject} by introducing variations in the dust foreground to investigate their effects on the analyses.

\section{Two Band Template Fit}
\label{sec:TFmethod}
We use a two band template fit method as a relatively simple method of probing $\beta_d$ variations across large portions of the sky.
This method has been used in prior \textit{Planck} foreground analyses to provide independent measures of the foreground spectral indices like $\beta_d$ to compare against their more involved foreground algorithms \citep{P2018_dust}. 
Although this template fit method is less flexible than the more complex multifrequency fitting analysis we also implement, it directly measures $\beta_d$ in different regions of the sky and helps set expectations for the level of signals we expect to see.

In the template fit method, the scaling of the dust from the template's frequency to a map at some other frequency can be converted to a measure of $\beta_d$ between the two frequencies.
The template fit method can become complicated as the number of components in the maps increases, since each component can scale differently between the template and the map under consideration. 
However, at the highest frequencies we consider, 353 GHz and 217 GHz, the maps are dominated by dust and, to a lesser extent, the CMB.
By subtracting an estimate of the CMB beforehand, we are left with maps dominated by dust emission, which can be used to derive $\beta_d$ between 353 GHz and 217 GHz.

We perform this analysis via minimizing the following $\chi^2$ in the $Q$ and $U$ maps for the scaling factor, $\alpha$, and offsets, $a$ and $b$:
\begin{align}
    \chi^2 &= \sum_{k}^{N_{pix}}(Q_{[217-CMB]}(k) - \alpha Q_{[353-CMB]}(k) - a)^2 \nonumber\\
    &+ (U_{[217-CMB]}(k) - \alpha U_{[353-CMB]}(k) - b)^2.\label{eqn:cc_chi2}
\end{align}

\noindent{The template scaling factor $\alpha$ can be converted to $\beta_d$ by setting $\nu_d$ = 353 GHz and $\nu$ = 217 GHz in the following relation, derived from Equation~\ref{eqn:model}},
\begin{equation}
    \alpha = \frac{g_{d}(\nu)}{g_{d}(\nu_d)}\left(\frac{\nu}{\nu_{d}}\right)^{\beta_d+1}\frac{\exp{(\frac{h\nu_d}{k_BT_d}}) -1}{\exp({\frac{h\nu}{k_BT_d}}) -1}
\end{equation}
and solving for $\beta_d$.

We restrict the above analysis to the three large and disjoint regions of the sky, as defined in Section~\ref{ssec:masks} and Figure~{\ref{fig:masks}}, to maximize the signal to noise ratio and determine if there is a significant variation in $\beta_d$ between these regions.

\section{Template Fit Results}
\label{sec:TFresults}
The template fit analysis requires the CMB to be removed from the frequency maps beforehand.
As described in Section~\ref{ssec:addprod}, we use the four independently derived CMB map estimates provided in the PR3 data release: those from the Commander, SEVEM, SMICA, and GNILC algorithms.

\begin{table}[h!]
\centering
\caption{$\beta_d$ as derived from template fitting}
\label{tab:TFres}
\begin{tabular}{c c c c} 
 \hline
 CMB estimate & Mask H    & Mask M     & Mask L \\ [0.5ex] 
 \hline\hline
 Commander & $1.57\pm0.07$\footnote{Uncertainties are from the standard deviation of 300 $\beta_d$ values recovered in identical analyses of the 300 baseline simulations, with the CMB perfectly removed.} & $1.54\pm0.03$ & $1.54\pm0.02$ \\ 
 SEVEM & $1.59\pm0.07$ & $1.55\pm0.03$ & $1.53\pm0.02$ \\
 SMICA & $1.53\pm0.07$ & $1.52\pm0.03$ & $1.52\pm0.02$ \\
 GNILC & $1.50\pm0.07$ & $1.51\pm0.03$ & $1.51\pm0.02$ \\ [1ex] 
 \hline
\end{tabular}
\end{table}

Table~\ref{tab:TFres} contains the results of the template fit analysis conducted over each of the three masked regions for each of the four CMB maps.
To assign uncertainties to the $\beta_d$ estimates, we repeat the template fit analysis on each of 300 baseline simulations, assuming perfect CMB removal, and take the standard deviation of $\beta_d$ over the simulations.

We find that across regions and CMB versions, these recovered $\beta_d$ values are consistent.
These values are also consistent with the overall mean and standard error $\beta_d$ value of $1.55\pm0.05$ recovered by \textit{Planck} over the full sky \citep{P2018_foregrounds}.

These results also suggest that it will be difficult to directly measure a variation in $\beta_d$ on the sky using the current data.
The template fit analysis should let us discern large-scale region spectral differences if there are significant variations between sky regions.
Further, switching the CMB estimate used has a comparable or even greater impact on $\beta_d$ values than switching the sky region.
We also checked the impact of fixing the offsets, $a$ and $b$, to zero.
Doing so shifts the derived $\beta_d$ by at most 0.01, or less than $0.5\sigma$ in the strictest case in mask L, and less than the shift due to switching the CMB estimate.

We also repeated this analysis on coarser 2-way partitions of the sky such as a north and south split and an east and west split, while keeping the plane masked.
No significant variations in $\beta_d$ were found.

Based on the results of this template fit analysis, rather than attempting to detect $\beta_d$ variations on the sky, we focus the more complex multifrequency fitting analysis on extracting an upper bound on $\beta_d$ variations and testing candidate $\beta_d$ models.

\section{Multifrequency Fitting}
\label{sec:fitting_method}
The template fit method is useful for directly computing $\beta_d$ on predetermined regions of the sky; however, it cannot detect or constrain variations in $\beta_d$ if lines of sight with different indices are assigned to the same region.
This can occur due to suboptimally drawing the borders of the analyzed regions or because the scale of $\beta_d$ variations is smaller than the size of the regions we consider.

Further, the template fit method also depends on the accuracy of the CMB signal removed from the 353~GHz and 217~GHz maps. 
As we discuss below in Section~\ref{ssec:priors}, the CMB maps provided by \textit{Planck} contain features related to residual systematics.
This may negatively affect the template fit analysis by introducing spurious correlations that scale in frequency like the CMB. 
Chance correlations between the dust and residual systematic signatures could shift the mean value of $\beta_d$.
They could also decrease the variance, but it is more likely that the presence of residual systematics would be uncorrelated and increase the $\beta_d$ variance. 

To address the concerns above, we implement a pixel-by-pixel multifrequency fitting analysis where we directly fit the parameters in Equation~\ref{eqn:model}.
This multifrequency fitting analysis is more flexible than the template fit analysis and allows for a test of the fidelity of the FFP10 noise simulations, and of the models at lower frequencies.
Lower frequencies are important for cosmology since they have higher CMB signal to noise ratios than 217 and 353~GHz.
Multifrequency fitting also allows us to introduce both random and systematic dust variations into the simulations to validate the methodology.

\subsection{Likelihood}
Given the number of bands available, it is possible to leave all or most of the parameters free in Equation~\ref{eqn:model} when fitting to the data.
However, there are known complications with fitting the spectral indices directly as they exhibit non-Gaussian behavior that require a careful treatment using priors \citep{2009_Dunkley, P2020}.
$T_d$ is often found to be somewhat degenerate with $\beta_d$, because of the formulation
of the dust SED function.
Further, as we are working primarily with the HFI bands, we do not expect to constrain $\beta_s$ or $A_{s, Q}$ and $A_{s, U}$.
Therefore, we perform simplified fits where we fix $\beta_s$, $\beta_d$, and $T_d$ to their best fit sky averages as determined by the \textit{Planck} collaboration.
We fix $\beta_s$ to -3.1 and $T_d$ to 19.6~K as in \cite{P2018_dust}.
We hold $A_{s, Q}$ and $A_{s, U}$ fixed to the 30~GHz map we use as the synchrotron template.
In our baseline fits, we also fix $\beta_d$ to 1.55 based on the earlier analysis in Section~\ref{sec:TFresults} and the constraints from \citet{P2018_foregrounds}.
We can assess the impact of spatial variations in $\beta_d$ by changing this fixed value or changing the input $\beta_d$ in the simulated maps and repeating the fit.

By holding the above parameters constant within each fit, and if the noise is Gaussian, the posterior distributions for the four remaining parameters, $A_{d,Q}$, $A_{d,U}$, $CMB_Q$, and $CMB_U$, will also be Gaussian.
These simplifications are justified as even if $\beta_d$ and $T_d$ are held fixed in any given fit, we can consider the impact of changing the assumed-per-fit or simulated value for $\beta_d$ on specific dust variation statistics, defined in Section~\ref{sec:fitting_results}, to investigate the spatial variations in the dust foreground.

For each pixel, the free parameters were fit by minimizing the negative log-posterior, which we define as an effective chi-squared, or $\chi^2_{eff}$, per pixel:
\begin{align}
    \chi^2_{\mathrm{eff}}& =\sum_{\nu} \left(d(\nu)_{[Q,U]} - m(\nu)_{[Q,U]}\right)C^{-1}\left(d(\nu)_{[Q,U]} - m(\nu)_{[Q,U]}\right)^{T} \nonumber\\
    &\quad + \sum_{\textrm{P}_i\in\textrm{Priors}} \ln(P_i).\label{eqn:eff_chi2}
\end{align}
Here, the first summation is over the four HFI frequency bands in the fit and corresponds to the usual $\chi^2$ with $d$ as the data, $m$ as the model as defined in Equation~\ref{eqn:model}, and $C$ as the QU covariance matrix for the noise in the pixel.
The second summation represents the prior on $CMB_Q$ and $CMB_U$, which is described below in Section~\ref{ssec:priors}.
Each quantity in Equation~\ref{eqn:eff_chi2} is evaluated independently per pixel.

\begin{figure*}[!ht]
\centering
\includegraphics[width=1.0\textwidth]{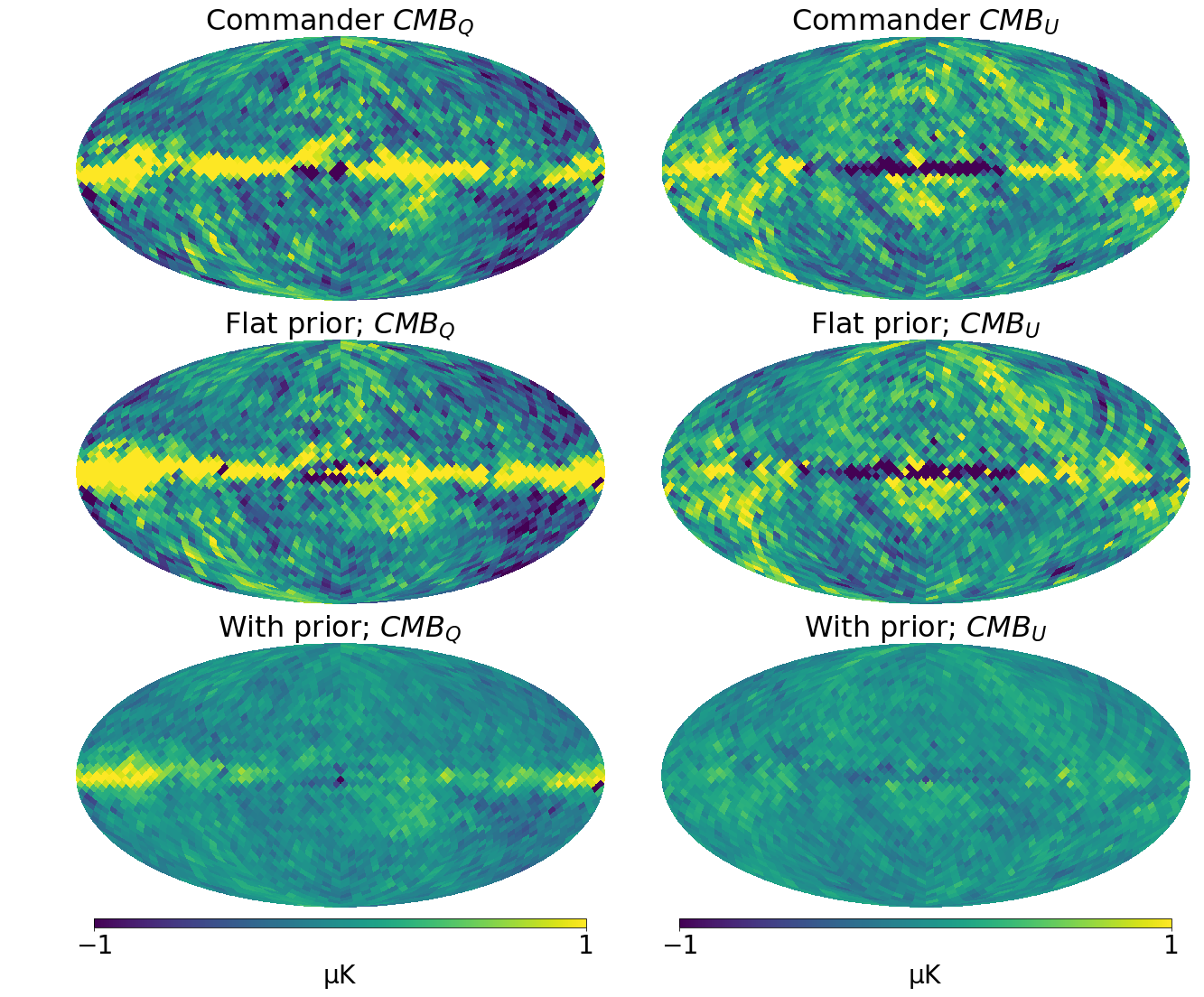}
\caption{{\it{Top row:}} CMB Q and U maps recovered by the \textit{Planck} Commander algorithm, which does not use
priors on the CMB pixel amplitudes. 
{\it{Middle row:}} CMB Q and U maps recovered 
by the fitting analysis described in Section~\ref{sec:fitting_method}, without priors.
{\it{Bottom row:}} As with the middle row, but with a Gaussian prior
on the CMB pixel amplitudes, based on the 2015F(CHM) (Plik) cosmology \cite{P2015_results}. There are large excursions in the CMB maps recovered by Commander and the no-prior analyses that trace out the \textit{Planck} scan pattern. Similar features are also present in the other three \textit{Planck} CMB maps generated by the \textit{Planck} GNILC, SEVEM, and SMICA analyses. As can readily be seen, the use of a prior substantially reduces these systematic features in the recovered CMB maps, largely confining them instead to 
the foreground components of the fit.}
\label{fig:cmb_w_wo_prior}
\end{figure*}

\subsection{Priors}
\label{ssec:priors}
Although the simplifications described above resolve the most complicated fitting issues when defining priors, we find that applying flat priors to all the remaining free parameters produces CMB map fits that contain recognizable systematics.  
To address this, we apply a Gaussian prior on the CMB pixel amplitudes that improves the recovery of the CMB in the fits.

The Gaussian CMB prior is defined by computing the covariance matrix in harmonic space for the same set of cosmological parameters used to generate the CMB for the simulations, with $10^9 A_s e^{-2\tau} = 1.881$, $\tau = 0.06$, and all other parameters set to those of the 2015F(CHM) (Plik) cosmology in table 6 of \cite{P2015_results}.
The harmonic space CMB covariance was converted to pixel space via a spherical harmonic transform operator.
Although the full CMB pixel space covariance matrix is singular, it suffices for defining the CMB prior, as the fits are conducted one pixel at a time. 

Figure~\ref{fig:cmb_w_wo_prior} shows examples of the CMB map recovered from the data by Commander, which does not use a CMB prior, and the analysis in this paper, with and without this CMB prior.
We find that without the CMB prior, there are large excursions from the input CMB realization that appear to be known systematic effects that trace the \textit{Planck} scan pattern.
These systematic patterns present in the \textit{Planck} data and CMB maps must be accounted for in any analysis.
We minimize such systematic excursions by applying a Gaussian prior to improve the performance of the foreground fit.
This means that residual systematics are primarily confined within the foreground parameters, which translates to a high dependence on the fidelity of the noise model used in the likelihood and in the simulations we calibrate against.

Use of the CMB prior differentiates the multifrequency analysis from that of e.g. Commander.
With the methodology established in this section, we proceed to implement it and show results in Section~\ref{sec:fitting_results}.

\section{Multifrequency Fitting Results}
\label{sec:fitting_results}
\begin{figure*}[t!]
\centering
\includegraphics[width=0.9\textwidth]{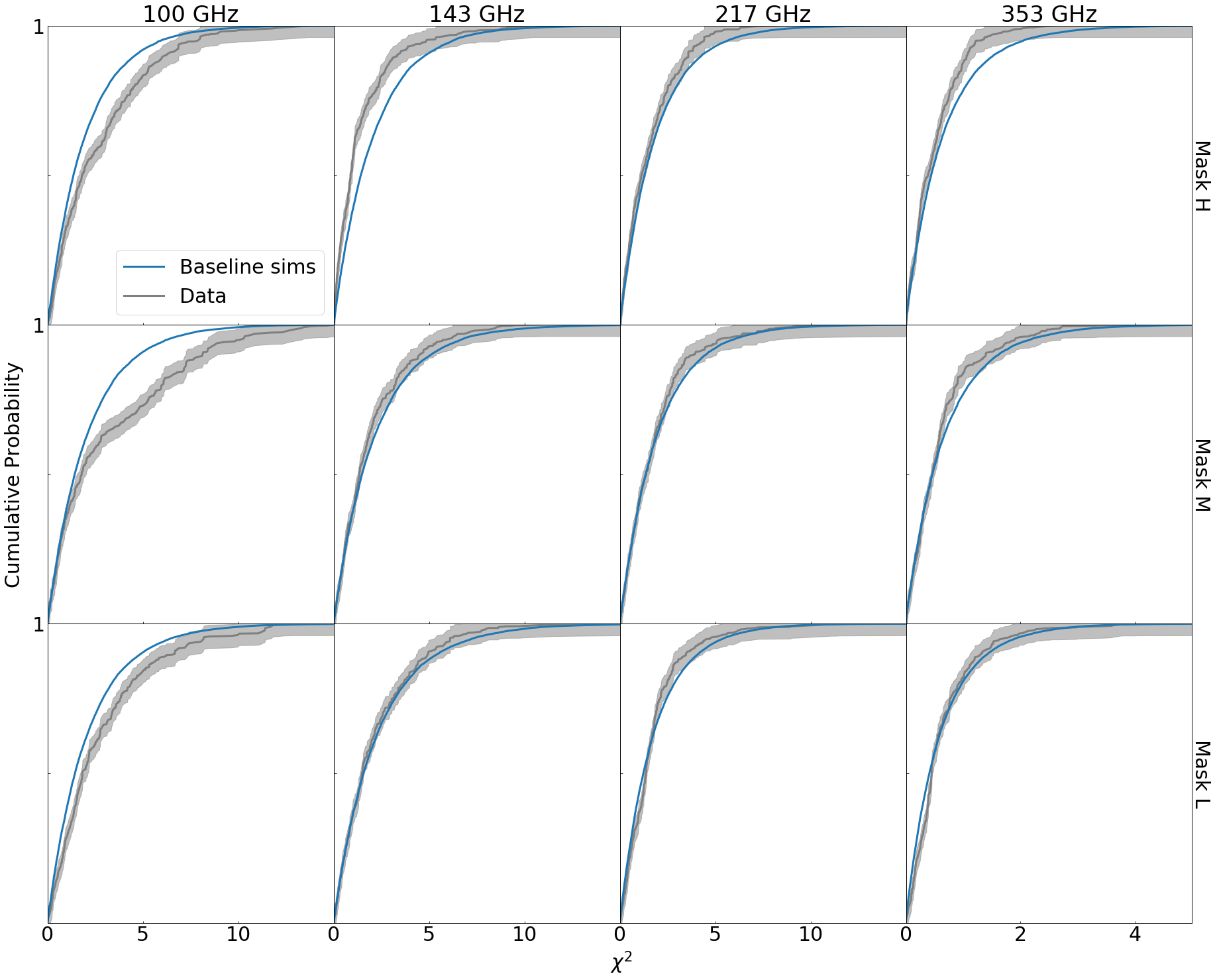}
\caption{$\chi^2$ empirical cumulative distribution functions (CDFs) from the multifrequency fits for the data and for all 300 baseline simulations together broken up into the four frequency bands and three sky regions we consider. The blue curve represents the empirical CDF from all 300 simulations. For a given value of $\chi^2$, the corresponding value of the curve represents the fraction of pixels with smaller $\chi^2$ over all 300 simulations. For example, if the blue curve was 1 at $\chi^2 = 10$, this means no pixel out of all the simulations has a greater $\chi^2$. The grey curve and shaded region corresponds to the empirical CDF for the data and a 68\% confidence interval around it, estimated using the DKW inequality \citep{DKW}. If the data and simulations agree, the blue curve should be entirely contained within this shaded region 68\% of the time. The data and the simulations appear consistent in most panels, except at 100 GHz and for mask H at 143 GHz.}
\label{fig:dvs_res}
\end{figure*}

The ultimate objective is to detect or constrain possible dust variations over the sky through probes of the spatial variation of $\beta_d$. 
This is straightforward for the two-band template fit analysis described above in Section~\ref{sec:TFresults}, as we solve for a constant $\beta_d$ over defined sky regions, but is more complicated for the multifrequency fitting analysis we now consider. Here, reliance on statistical properties inferred from multifrequency sky simulations establishes dust variation constraints.

Constraints on variations in $\beta_d$ over the sky are highly sensitive to the accuracy of the noise model and possible deviations from it in the form of genuine foreground variations or unaccounted-for systematics in the data that are not included in the simulations.
In Section~\ref{ssec:dvs_comps}, we investigate how well the full-sky
\textit{Planck}-like simulations, in which 
we specify the CMB and foreground components that we add
to the FFP10 \textit{Planck} noise and residual systematic simulations, capture the behavior of the data.

We present results of the investigations of the dust properties in the data in \ref{ssec:bd_offset}, \ref{ssec:bdvary}, and \ref{ssec:bd_inject}.
In these sections, we make use of two statistical methods to capture the signal due to a spatially varying dust foreground:
\begin{enumerate}
  \item We examine the total $\chi^2$ distribution and the $\chi^2$ distributions per mask region per band. 
A spatial variation in $\beta_d$ should shift the $\chi^2$ distribution relative to the simulations where the $\beta_d$ is constant. 

  \item We estimate the dust contribution to the residual in each band by correlating the residual maps of the fit against the 353 GHz map we use as the dust template.
If there is spatial variation in $\beta_d$, and the fit assumes a spatially uniform $\beta_d$, the residual maps will contain an excess dust contribution, which would appear as a non-zero correlation with the dust template.  
However, additional spurious correlations between the dust template and fit residual maps may be introduced as a result of the fitting analysis.  
Therefore, we use simulations to estimate this spurious fitting-induced uncertainty.
\end{enumerate}

In subsections~\ref{ssec:bdvary} and \ref{ssec:bd_inject}, we also introduce simulations in which the foreground dust model has an intrinsic spatial variation.

\subsection{Comparisons between data and simulations}
\label{ssec:dvs_comps}

In this section, we work with the 300 baseline simulations generated using the prescription described in Section~\ref{ssec:sims}.
These simulations represent the baseline case where the foregrounds, noise, and systematics are known.
Any mismatch between these fiducial simulations and the actual data must be the result of either more complex foregrounds or unaccounted-for systematics in the data that are not included in the FFP10 simulations.

The $\chi^2$ distributions of the foreground fits performed on the data and the ensemble of simulations are one way of comparing the foreground fits for the simulations to those for the data.
The foreground fits produce $\chi^2$ values per pixel per band. 
From these $\chi^2$ values, we can compute band-by-band empirical cumulative distribution functions for the fits to the data and the simulations.
Figure~\ref{fig:dvs_res} shows these band-by-band empirical cumulative distribution functions for the $\chi^2$ in each of the masked regions estimated over all 300 simulations together and for the data.
In general, the data and simulations appear to agree, except for all masks at 100 GHz and mask H for 143 GHz.

To quantify the (dis)agreement between the data and simulation $\chi^2$ distributions in Figure~\ref{fig:dvs_res}, we estimate a numerical probability-to-exceed (PTE) based on the Kolmogorov–Smirnov (KS) test statistic.
We defined a reference $\chi^2$ distribution as the cumulative $\chi^2$ distribution over the 300 baseline simulations in each mask region in each band.
Then, we computed the KS test statistic between the $\chi^2$ distribution of each simulation and the reference $\chi^2$ distribution.
We also computed the KS test statistic between the $\chi^2$ distribution of the data and of the reference distribution.
By comparing the KS statistic computed in this way for the data against 300 KS statistics computed for the baseline simulations, we numerically estimate the PTE for the data by counting the number of simulations with KS statistic values greater than that of the data and dividing by the total number of simulations.
For example, in mask H at 100 GHz, we find 9 simulations with KS statistic values greater than that of the data from which we compute a PTE of 9/300.
The results of this procedure are shown in Table~\ref{tab:KStest_dvs}.

\begin{table}[h!]
\caption{Data PTE per band per mask}
\label{tab:KStest_dvs}
\centering

\begin{tabular}{c c c c} 
 \hline
  & \multicolumn{3}{c}{PTE (\# sims out of 300)} \\
 Band & Mask H & Mask M & Mask L \\
 \hline\hline
100 GHz & 9\footnote{number of simulations with KS statistic greater than that of the data.} & 18 & 113 \\
143 GHz & 8 & 104 & 175 \\
217 GHz & 115 & 287 & 135 \\ 
353 GHz & 56 & 70 & 165 \\
 \hline
\end{tabular}
\end{table}

Low PTE values for the data can be seen for masks M and H at 100 GHz and mask H at 143 GHz. 
This result is consistent with Figure~\ref{fig:dvs_res}. 
Although Figure~\ref{fig:dvs_res} suggests that the PTE value of the data for mask L at 100 GHz might be low, its PTE value ($\approx 0.38$) as computed in this test is not particularly low.  
Of the four cases discussed here, the KS statistic for mask L at 100 GHz is the least discrepant in Figure~\ref{fig:dvs_res}, which may explain why the PTE value is acceptable.

\begin{figure}[t]
\centering
\includegraphics[width=\linewidth]{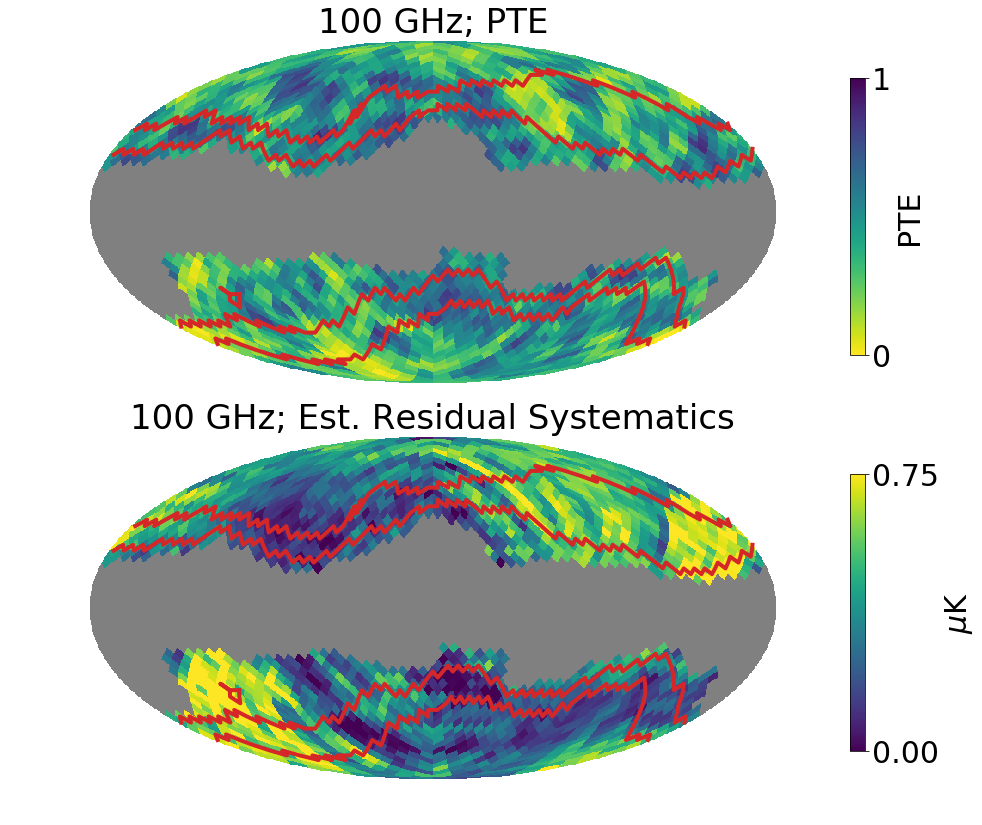}
\caption{Top plot, in Galactic coordinates, depicts the $\mathrm{PTE}$, smoothed to $5^{\circ}$ FWHM, as estimated in each pixel at 100 GHz by comparing the data $\chi^2$ against the distributions of $\chi^2$ over the 300 baseline simulations. Also in Galactic coordinates and at the same resolution, the bottom plot is an estimate of sky regions most affected by the potential presence of analog to digital non-linearity not fully accounted for in the FFP10 noise model (see text). The red contours denote the edges of the three masks. Comparing the top plot with the bottom plot, there are bright bands along the ecliptic plane which trace out the \textit{Planck} scan pattern in the top right and bottom left of both maps. These features could explain the excess in $\chi^2$ and discrepant PTEs at 100 GHz in Figure~\ref{fig:dvs_res} and Table~\ref{tab:KStest_dvs}.}
\label{fig:c2pte_perpix}
\end{figure}

\begin{figure*}[t]
\centering
\includegraphics[width=1.0\linewidth]{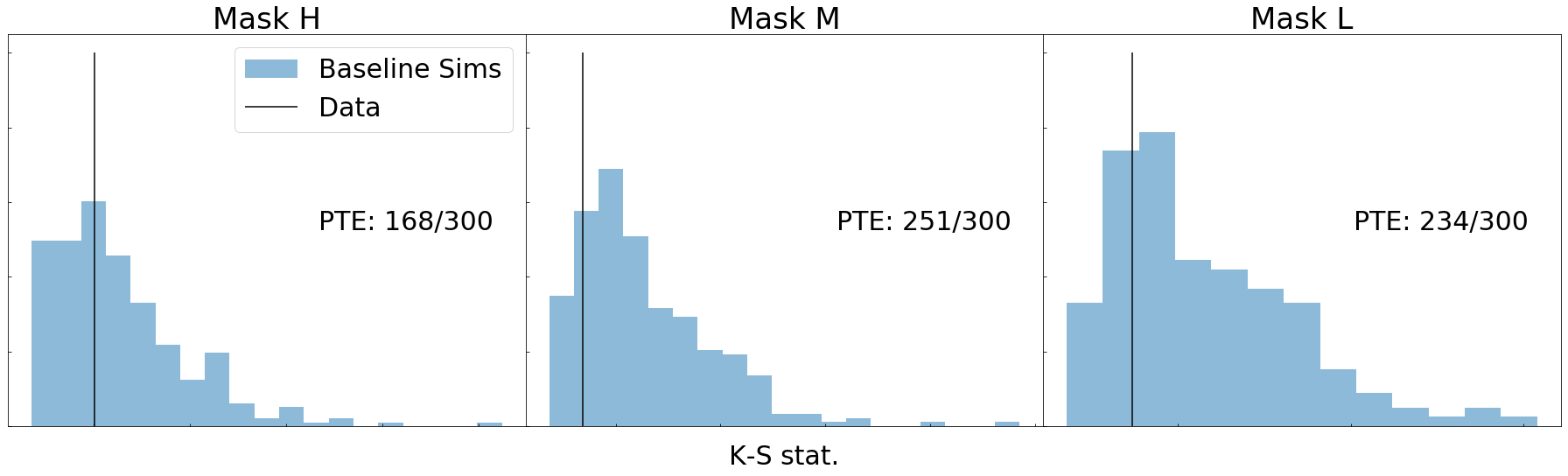}
\caption{Histograms of the KS statistic comparing the total $\chi^2$ distribution in each of the three masked regions per baseline simulation against the cumulative $\chi^2$ distribution over all 300 baseline simulations. We also plot the KS statistic value for the data compared against the same cumulative $\chi^2$ distribution over all 300 baseline simulations as a vertical black line. After estimating the PTE for the data from these histograms, by comparing the value for the data against the distribution of values from the simulations, we find that the data and baseline simulations are consistent in all three masks.}
\label{fig:KShist_tot_dvs}
\end{figure*}

We further consider the low PTE values shown in Table~\ref{tab:KStest_dvs} at 100 GHz by numerically estimating the PTE of the data $\chi^2$ per pixel and directly comparing to the $\chi^2$ distribution for each pixel over the 300 baseline simulations.
The PTE per pixel at 100 GHz is shown in Figure~\ref{fig:c2pte_perpix} alongside an estimate of residual unaccounted-for systematics in the 100 GHz map. 
This estimate of unaccounted-for systematics was produced by subtracting the SRoll2 100~GHz map and the mean of the 300 FFP10 noise simulations at 100~GHz from the PR3 100~GHz map we use.
As SRoll2 reduces the residual systematics, $\mathrm{PR3-SRoll2}$ yields an estimate of residual systematics present in the PR3 map. 
Note that SRoll2 was not used in the rest of the analysis due to reasons outlined in Section~\ref{ssec:data}.
We estimate the noise covariances used in the fitting analysis from the 300 FFP10 noise simulations.
Thus, further removing the mean of the 300 FFP10 noise simulations at 100~GHz to compute $\mathrm{PR3-SRoll2 - mean_{300 sims}}$ yields the estimate of the residual systematics unaccounted-for in the FFP10 simulations for Figure~\ref{fig:c2pte_perpix}. 
In Figure~\ref{fig:c2pte_perpix}, which is in Galactic coordinates, there are bright bands along the ecliptic plane in both maps which trace out the \textit{Planck} scan pattern in the top right and bottom left. 
These features could explain the excess in $\chi^2$ and discrepant PTEs at 100 GHz in Figure~\ref{fig:dvs_res} and Table~\ref{tab:KStest_dvs}.

Estimating a similar PTE per pixel at 143 GHz, we find extensive regions of low PTE in mask H, consistent with the lower than expected $\chi^2$ in Figure~\ref{fig:dvs_res} for mask H.
These problems at 100~GHz and 143~GHz are consistent with known \textit{Planck} uncertainties in simulating analog to digital converter non-linearity effects in the FFP10 simulations, with a known overestimate of this effect at 143~GHz \citep{P2018_HFI}.

As discussed above, statistical inconsistencies in some of the regions and bands may be caused by possible unaccounted-for systematic effects and mismatches in the noise model used in the FFP10 noise simulations versus the noise in the data.  
Spatial variations in the polarized synchrotron spectral index, $\beta_s$, could also play a partial role in the 100~GHz PTE results, as large variations have the potential to be competitive with the estimated magnitude of residual systematics in the 100~GHz map.  
As of this writing, however, there is insufficient observational data to sufficiently constrain these variations over the full sky.
The \textit{Planck} data lack the necessary signal to noise ratio for a statistically significant detection of $N_{\mathrm{side}}=16$ pixel variations at high latitude \citep{P2018_foregrounds}. 
Analyses combining 2.3~GHz S-PASS data with Planck and/or WMAP have sky coverage limitations and uncertainties associated with Faraday depolarization \citep[e.g.][]{Krachmalnicoff2018, Fuskeland2021}.

However, if we consider a similar KS-based comparison computed using the total $\chi^2$ distribution, as in Figure~\ref{fig:KShist_tot_dvs}, instead of breaking the $\chi^2$ up band-by-band, we find closer agreement between the data and simulations.
Numerically estimating the PTE of the data from the histograms in Figure~\ref{fig:KShist_tot_dvs}, we find that 168 of the 300 simulations for mask H have a KS statistic value greater than or equal to that of the data.
We represent this PTE as 168/300.
We find PTE values of 251/300 in mask M, and 234/300 in mask L.
The total $\chi^2$ distribution is less affected by problems in individual bands or regions, which we take advantage of in the later analyses.

\begin{figure*}[t]
\centering
\includegraphics[width=0.9\linewidth]{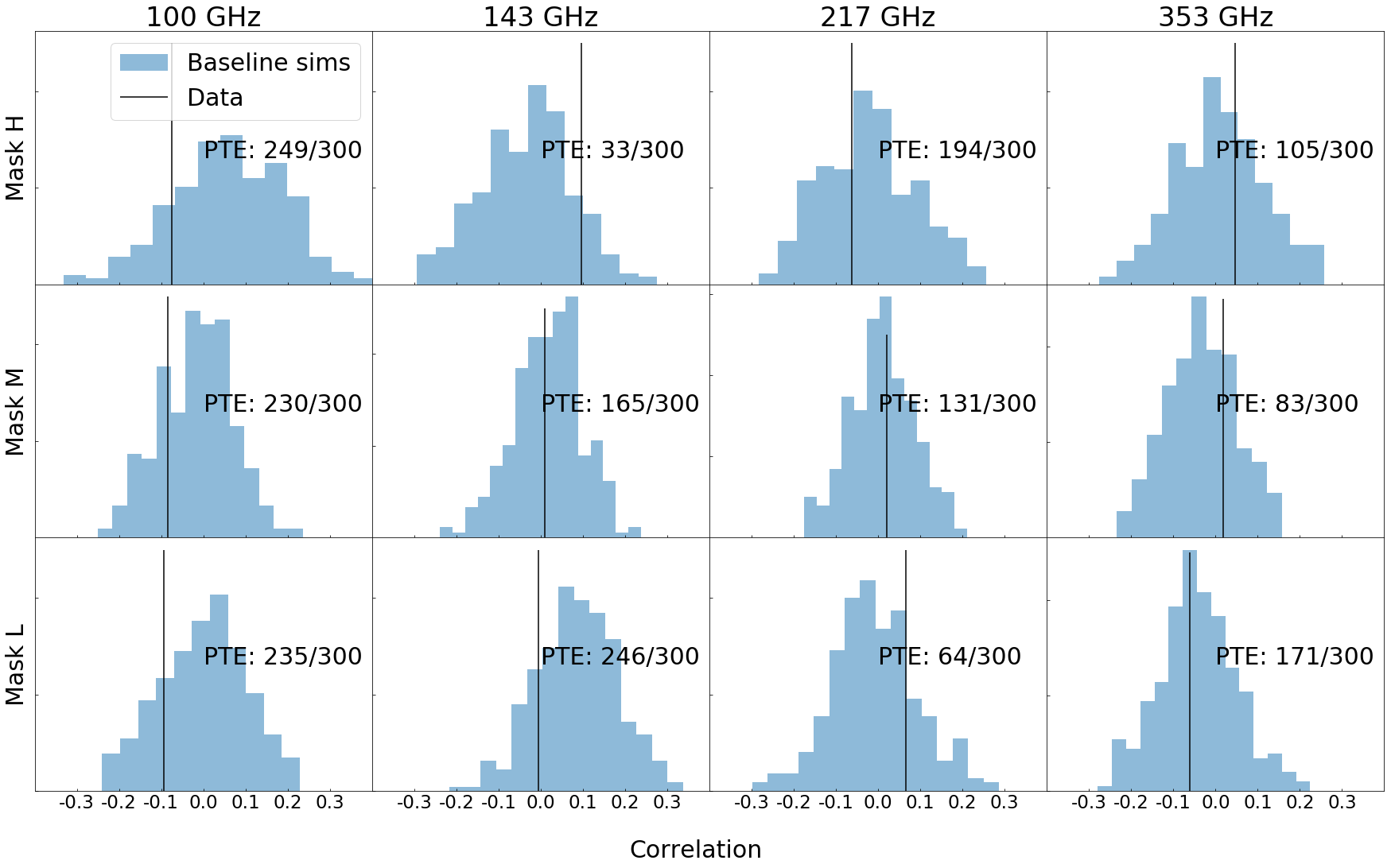}
\caption{Residual-dust correlations for the data and the ensemble of 300 baseline simulations for each masked region and frequency band. The correlation for the data is represented by a vertical black line and the distribution from the simulations is plotted as the blue histogram. These values are computed by correlating the residuals of the multifrequency fits in each region against the corresponding regions in either the data or simulated 353 GHz map. The data and simulation ensemble appear to be consistent in every panel. We also include the numerical estimate of the PTE from comparing the value for the data against the distribution of values from the simulations.}
\label{fig:dvs_cor}
\end{figure*}

In addition to considering the $\chi^2$ distributions, we can look at the correlation of the residual maps against the 353 GHz map, which is dominated by dust and can be used as a proxy for the dust signal.
By correlating the residuals against the dust signal, this approach focuses on the morphology of the dust foreground. 

Figure~\ref{fig:dvs_cor} shows the result of this residual-dust correlation procedure for the data and the simulations.
Here, for all bands and masked regions, the data fall within the spread in the simulations and have consistent PTEs. 

The two methods of comparing the data and the simulations presented here can both capture mismatches between the data and simulations, but have different sensitivity to different effects.
The $\chi^2$ distributions are more sensitive to outliers in the frequency maps while the correlation based approach is more sensitive to spatial structure in the frequency maps.
Thus, it is not surprising that the data and simulations are more discrepant in one comparison than in the other.
It appears that although there are residual systematics in the 100 GHz and 143 GHz frequency maps, which cause disagreement in some regions for the band-by-band KS statistic based comparison, they are not strongly correlated with the dust and average out when looking at the total $\chi^2$ distributions.

\subsection{Offset in \texorpdfstring{$\beta_d$}{dust beta}}
\label{ssec:bd_offset}
\begin{figure*}[t]
\centering
\includegraphics[width=0.9\linewidth]{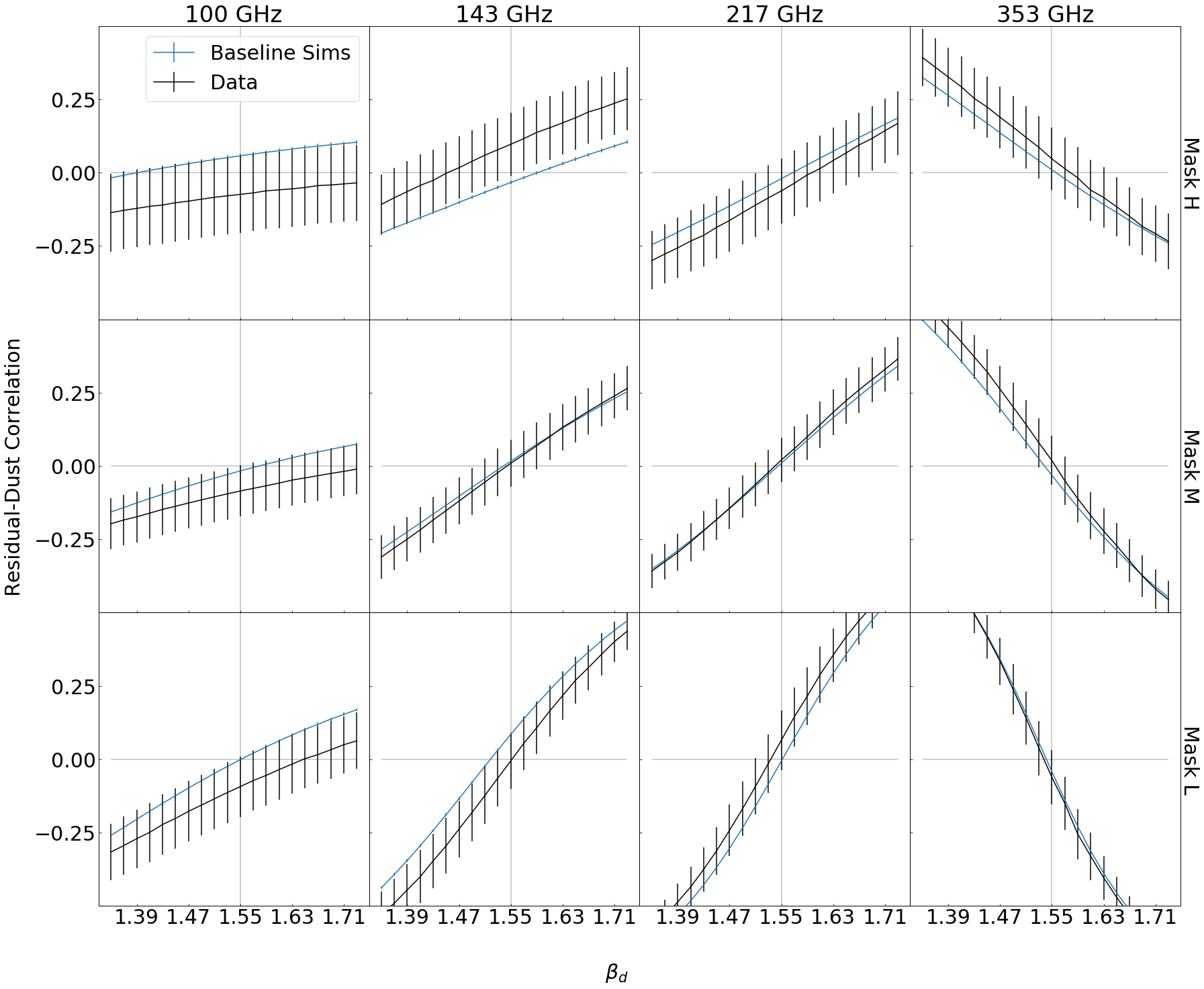}
\caption{Change in the residual-dust correlations as the assumed $\beta_d$ in the fit is stepped from 1.35 to 1.75 while the simulated $\beta_d$ is held fixed at 1.55. The values and uncertainties for the simulations are taken from the mean and standard error of the mean correlation of the simulation ensemble. The errors on the data are taken from the standard deviation of the correlations for the simulations. The gray horizontal line represents 0 correlation, which is expected for a perfect foreground removal. The gray vertical line is at $\beta_d = 1.55$, which is the input value for the simulations and roughly the value expected from \textit{Planck} constraints on the mean $\beta_d$. In most panels, both the data and simulations are consistent with each other and with $\beta_d = 1.55$. We attribute instances where the simulations are offset from 0 correlation at $\beta_d = 1.55$ to the structure of the noise in the FFP10 noise simulations and the necessity of using the simulated 353 GHz map, which includes the CMB and noise, in the correlations to match the treatment of the data. Repeating this exercise with white noise simulations and correlating against the simulation input dust template at 353 GHz eliminates all offsets from 0 correlation at $\beta_d = 1.55$.}
\label{fig:bdstep_cor}
\end{figure*}

We now investigate any overall difference between the fit $\beta_d$ and the simulated or true $\beta_d$.
Working with the same 300 baseline simulations from Section~\ref{ssec:dvs_comps} where the simulated $\beta_d$ is 1.55, we step through the $\beta_d$ assumed in the dust fit from 1.35 to 1.75 in steps of 0.02. 

Figure~\ref{fig:bdstep_cor} shows this effect in the residual-dust correlations for the simulations and the data.
In most panels, the residual-dust correlations for the simulations are consistent with no correlation with dust at the input $\beta_d = 1.55$ and shift further from zero the further the $\beta_d$ used in the foreground fit is from 1.55.
Some of the panels, which include mask H at 100 GHz and 143 GHz, show offsets from from 0 correlation at $\beta_d = 1.55$.
We attribute these offsets to the structure of the noise in the FFP10 noise simulations we use in the baseline simulations and the necessity of using the simulated 353 GHz map, which includes the CMB and noise, in the correlations to match the treatment of the data.
Repeating this analysis with white noise in the simulations and correlating against the simulation input dust template at 353 GHz eliminates all such offsets from 0 correlation at $\beta_d = 1.55$.

A similar trend holds for the data; they prefer a value near $1.55 \pm 0.02$ in most panels.
This matches the \textit{Planck} collaboration's finding of a central value of $\sim1.55$, depending on the choice of analysis, for the full-sky $\beta_d$ distribution \citep{P2018_foregrounds}.
This is also consistent with the results of the template fit analysis in Section~\ref{sec:TFresults}.
Further, the overall behavior of the data as we step through $\beta_d$ is consistent with the spread in the simulations.
This shows another area of agreement between the data and simulations and between this analysis and prior analyses.

Considering Figure~\ref{fig:bdstep_cor} further, the data (and simulations) are more constraining with steeper changes in the correlations as we move to lower masks and thus higher signal to noise ratio for the dust.
Also, 217 GHz and 353 GHz appear more constraining than the other 2 bands, which is not surprising as they are dominated by dust.

\subsection{Gaussian Variations in \texorpdfstring{$\beta_d$}{dust beta}}
\label{ssec:bdvary}
\begin{figure}[t]
\centering
\includegraphics[width=1.0\linewidth]{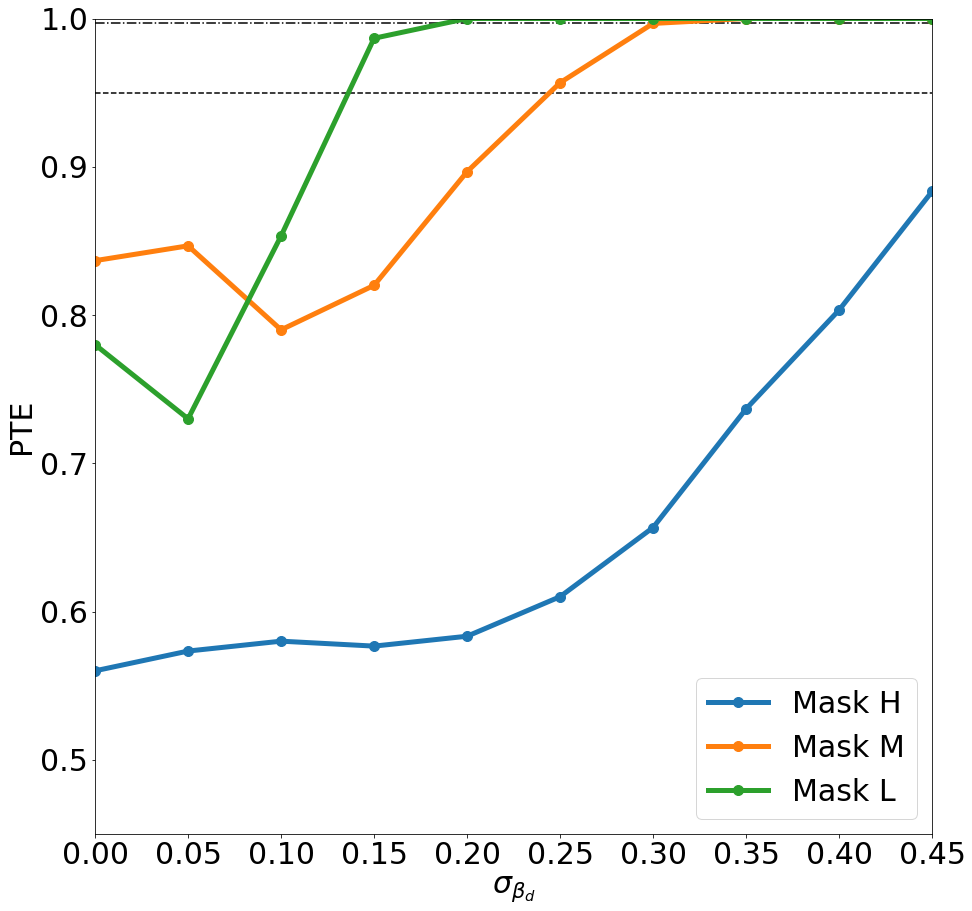}
\caption{Change in the estimated PTE of the data compared against simulations with an increasing injected Gaussian variation in $\beta_d$ of $\sigma_{\beta_d}$ for each masked region. The dashed horizontal line represents a PTE of 0.95 and the dash-dotted line represents a PTE of 0.997. We can rule out $\sigma_{\beta_d} = 0.15$ in mask L and $\sigma_{\beta_d} = 0.30$ in mask M. Mask H is the least constraining and we cannot significantly rule out a level of $\sigma_{\beta_d}$ in the range we test. This decrease in constraining power as higher latitudes are considered is consistent with the dust signal becoming fainter as one moves away from the plane.}
\label{fig:KStest_bdvary}
\end{figure}

Above, we found that the overall $\beta_d$ derived from the data using multifrequency fitting analysis is consistent with expectations.
Now, as a simple model of spatial variation in $\beta_d$ around $1.55$, we test a random Gaussian variation and set limits on the allowed $\sigma_{\beta_d}$.

To do this, rather than modifying the $\beta_d$ assumed in the fit as before, we produce ten sets of 300 simulations by injecting a Gaussian variation defined by $\sigma_{\beta_d}$ into the uniform $\beta_d=1.55$ map used in the simulations.
We increased $\sigma_{\beta_d}$ from 0 to 0.45 in ten steps of 0.05, which defines each set of 300 simulations.
For each simulation, we ran the fitting analysis with $\beta_d = 1.55$ fixed in the fits, and analyzed the resulting $\chi^2$ distributions.
We do not show results from the residual-dust correlations, as they are less informative for probing $\sigma_{\beta_d}$ since injecting a random variation into the simulated dust foreground should not affect the overall correlation of the residuals and the dust signal. 

To set upper bounds on the observed level of $\sigma_{\beta_d}$, we use an approach similar to the KS based comparisons in Section~\ref{ssec:dvs_comps}. 
We estimate a PTE for the data compared to each set of 300 simulations for each of the ten $\sigma_{\beta_d}$ values.
Here, we expect the PTE values to increase to 1 as the injected $\boldsymbol{\sigma_{\beta_d}}$ increases.
This is because greater $\sigma_{\beta_d}$ variations in the simulations will increase the $\chi^2$ from the fits to the simulations. 
As the simulations are always compared to the same fit to the data, the $\chi^2$ distributions from increasing $\sigma_{\beta_d}$ in the simulations will gradually shift to higher values with respect to the data, resulting in a larger PTE when comparing the data to the simulations.
This is opposite the standard expectation of smaller PTEs when determining upper bounds using analyses which compare a changing data model to a fixed distribution from simulations.

Figure~\ref{fig:KStest_bdvary} depicts the results of this procedure. 
This analysis sets upper bounds of $\sigma_{\beta_d} = 0.15$ with a PTE = 296/300 in mask L and $\sigma_{\beta_d} = 0.3$ with a PTE = 299/300 in mask M.
Mask H is the least constraining region and we cannot set an upper bound within the range of $\sigma_{\beta_d}$ we test, consistent with having the weakest dust signal.
For the highest $\sigma_{\beta_d}$ we test in mask H, 0.45, we find a PTE of 264/300. 

\citet{P2018_dust} estimates $\sigma_{\beta_d}=0.092\pm0.005$ via the power spectra of six nested regions on the sky with effective sky fractions of 24 to 71\%.
This constraint is a standard deviation over highly correlated $\beta_d$ measurements for these six nested regions, so it is natural that this $\sigma_{\beta_d}$ constraint is tighter than the one derived here.
Other differences may arise due to the differences between the pixel-based analysis above which focuses on effects at a pixel scale of about $3.7^{\circ}$ and a power spectrum based approach which considers a wide range of angular scales.

HFI instrument noise and systematics are important limiting factors in the estimation of $\sigma_{\beta_d}$. 
By downgrading the map resolution further, one might hope to achieve greater sensitivity and tighten constraints on $\sigma_{\beta_d}$, although with the loss of smaller scale information.
However, the constraints will only tighten at lower resolutions if the noise and other components in the maps integrate down faster than the signal that is due to spatial variations in $\beta_d$.

In alternative simulations with white noise where the signal due to spatial variations in $\beta_d$ is allowed to dominate at large scales, we confirm that the constraining power improves at lower resolutions. 
However, there are large scale noise and systematic structures in the data-like \textit{Planck} FFP10 simulations that dominate at lower resolutions over the Gaussian variations we inject. 
For this reason, we find that constraints at $N_{\mathrm{side}}=8$ do not tighten compared to those we derive at $N_{\mathrm{side}}=16$. 
In a similar vein, dust analyses and constraints from higher resolution analyses like those performed at $N_{\mathrm{side}}=128$ by \citet{PySM2017}, which found $\sigma_{\beta_d}$ in the 0.2–0.3 range, and at $N_{\mathrm{side}}=2048$ by \citet{P2018_foregrounds} cannot be reliably extrapolated to e.g. $N_{\mathrm{side}}=16$ by assuming a statistical $\sqrt{N_{\mathrm{pix}}}$ dependence. 
Each constraint must be interpreted in the context of the analysis resolution. 
Ultimately, the analysis presented here shows that at $N_{\mathrm{side}}=16$ the Planck data do not rule out large (> 0.1) $\sigma_{\beta_d}$ values, especially at higher latitudes where the dust signal is faint.

We next examine the impact of these constraints on upcoming high sensitivity experiments, using \textit{LiteBIRD} as an example.  
We generate 300 simulations at $N_{\mathrm{side}} = 16$ where simulated \textit{Planck}-like 353~GHz maps were used to clean simulated \textit{LiteBIRD}-like 140 GHz maps, which is the designated cosmology band. 
The dust amplitude was set to the \textit{Planck} 353~GHz map and scaled down to 140~GHz using either a spatially constant $\beta_d = 1.55$ or with a spatially varied Gaussian $\sigma_{\beta_d} = 0.15$, $0.3$, or $0.45$ injected per mask regions L, M, and H.
Noise was added to the 353~GHz simulated maps using the same FFP10 noise simulations used in our fitting analysis.
\textit{LiteBIRD} white noise was added at 140~GHz, based on the level specified in \citet{Hazumi2020} Table 3. 
A CMB realization was also added to each simulated map, assuming the same fiducial cosmology described in Section~\ref{ssec:sims}.

\begin{figure}[t]
\centering
\includegraphics[width=1.0\linewidth]{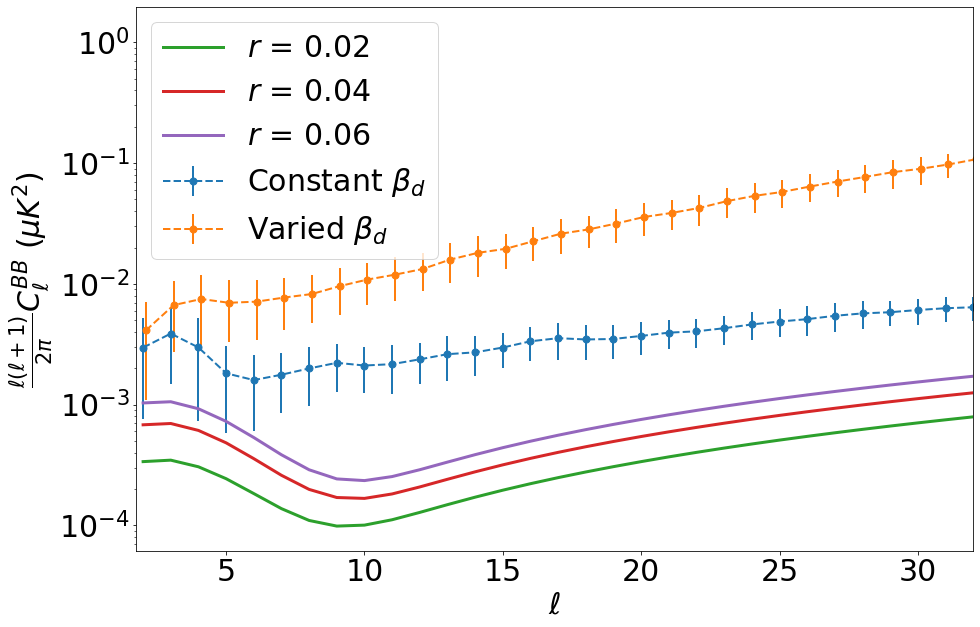}
\caption{Mean and standard deviation over 300 simulations for B-mode power spectra computed from residual maps after template cleaning dust+CMB+noise simulations of \textit{LiteBIRD} 140 GHz using \textit{Planck} 353 GHz (see text). These spectra were computed on the union of the L, M, and H mask regions using \texttt{PolSpice}. Theory CMB spectra for three levels of the tensor-to-scalar ratio, $r$, are shown for reference. The bump at low-$\ell$ for the residual spectra is due to the structure of the \textit{Planck} noise in the 353 GHz FFP10 noise simulations.}
\label{fig:clean140spectrum}
\end{figure}

The simulated 140~GHz maps were cleaned using the simulated 353~GHz maps via the same template cleaning analysis described in Section~\ref{sec:TFmethod}, with the fit restricted to the 60\% of the sky in the union of the three L, M, and H mask regions.
We then subtracted the input CMB from the recovered CMB to produce residual maps.
\texttt{PolSpice} \citep{Polspice2004} was used to compute the B-mode power spectrum of these residual maps on the union of the three mask regions. 
Figure~\ref{fig:clean140spectrum} shows the mean and standard deviation of these residual power spectra for the 300 simulations in the spatially constant and varied $\beta_d$ cases.
For reference, the figure also includes theory spectra for three values of the tensor-to-scalar ratio, $r = 0.06,$ $0.04,$ and $0.02$.
The current upper limit using data from a combination of experiments including BICEP2/Keck and \textit{Planck} is $r < 0.044$ (95\% CL) \citep{Tristam2021, BICEP2018}.

The orange spectrum in Figure~\ref{fig:clean140spectrum} is derived from template-cleaned residuals in the simulated case where the maximum allowed variation in the dust spectral index is assumed.  
This spectrum represents an order of magnitude increase in residual power in the template cleaned 140 GHz map compared to the spatially constant case shown in blue.
Although even a spatially constant $\beta_d$ is not sufficient to allow a \textit{Planck}-like 353 GHz polarization map to be used in future B-mode searches, the effect of these levels of $\sigma_{\beta_d}$ should be accounted for by future experiments.
Relying on dust model predictions based on the assumption that $\beta_d$ is constant or only slightly varies over the high latitude sky may be overly optimistic.

\subsection{Injecting external \texorpdfstring{$\beta_d$}{dust beta} maps}
\label{ssec:bd_inject}
\begin{figure}[t]
\centering
\includegraphics[width=1.0\linewidth]{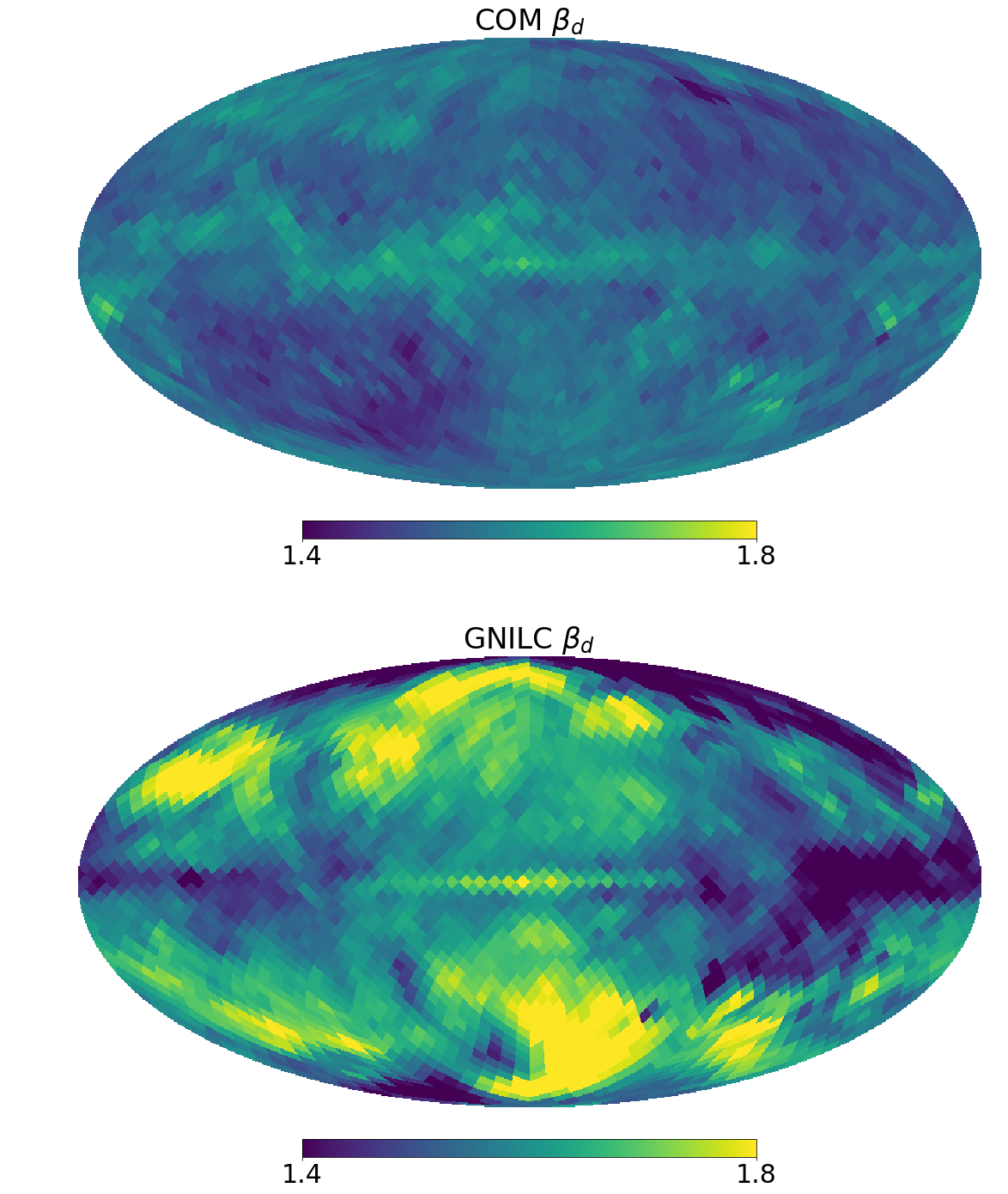}
\caption{$\beta_d$ maps produced by the 2015 Commander (top) and GNILC (bottom) foreground analyses. Note the significant differences in structure and amplitude of variation. In Section~\ref{ssec:bd_inject}, we inject these dust signatures into simulations and find that differences of this level can be detected via the analysis.}
\label{fig:com_gnilc_bd_maps}
\end{figure}

Instead of injecting random variations in ${\beta_d}$ per pixel, we now introduce large scale coherent spatial variations by using $\beta_d$ and $T_d$ maps produced by the Commander and GNILC foreground separation algorithms and released by the \textit{Planck} Collaboration \citep{P2015_foregrounds}.
These maps were released as part of the 2015 PR2 data release and take advantage of the higher signal to noise ratio in the intensity data to make a measurement of $\beta_d$ and $T_d$ per pixel.
Although the data and simulations we use are part of the 2018 PR3 data release, similar paired $\beta_d$ and $T_d$ maps for the PR3 dataset were not available.

Figure~\ref{fig:com_gnilc_bd_maps} shows the $\beta_d$ maps we use, downgraded to the $3.7^{\circ}$ analysis resolution.
There are significant differences in the structure and amplitude of variation between the Commander and GNILC $\beta_d$ maps.
We perform a similar downgrading of the $T_d$ maps and use these external  maps to define the dust component in two new sets of 300 simulations, one for the Commander model and another for the GNILC model.

\begin{table}[h!]
\centering
\caption{PTEs for each configuration of the multifrequency fitting}
\label{tab:PTEsummary}
\begin{tabular}{c c c c} 
 \hline
  & \multicolumn{3}{c}{PTE (\# sims out of 300)} \\ Configuration & Mask H & Mask M & Mask L \\
 \hline\hline
 KS test; baseline sims & 168\footnote{Number of simulations with KS statistic or correlation greater than that of the data.} & 251 & 234\\ 
 Res.-Dust\footnote{Residual-Dust correlation (see Section~\ref{ssec:dvs_comps})}; 100 GHz & 249 & 230 & 235 \\
 Res.-Dust; 143 GHz & 33 & 165 & 246 \\
 Res.-Dust; 217 GHz & 194 & 131 & 64 \\ 
 Res.-Dust; 353 GHz & 105 & 83 & 171 \\
 \hline\hline
 KS test; GNILC sims & 198 & 90 & 78 \\
 Res.-Dust; 100 GHz & 255 & 213 & 214 \\
 Res.-Dust; 143 GHz & 9 & 3 & 36 \\
 Res.-Dust; 217 GHz & 66 & 1 & 1 \\
 Res.-Dust; 353 GHz & 263 & 300 & 300 \\
 \hline\hline
 KS test; Commander sims & 121 & 269 & 125 \\
 Res.-Dust; 100 GHz & 265 & 267 & 298 \\
 Res.-Dust; 143 GHz & 42 & 277 & 300 \\
 Res.-Dust; 217 GHz & 190 & 283 & 300 \\
 Res.-Dust; 353 GHz & 112 & 1 & 1 \\
 \hline
\end{tabular}
\end{table}

To investigate whether the data prefer the baseline, Commander, or GNILC dust model, we performed two additional sets of fits where we fixed both $\beta_d$ and $T_d$ to the input spatially varying Commander or GNILC model when fitting these two new sets of simulations.
This was repeated for the data.
This results in two new sets of fits for data and simulations, one for each dust model.
As in the baseline case, we consider both the KS-based $\chi^2$ comparison and the residual-dust correlations.

Table~\ref{tab:PTEsummary} includes the PTEs for these KS comparisons.
Comparing the KS-based PTEs for the GNILC or Commander dust models to those for the baseline case, we conclude that the KS-based comparison cannot significantly distinguish between the baseline, GNILC, or Commander dust models.

Table~\ref{tab:PTEsummary} also includes the PTEs from the residual-dust correlation distributions for these fits.
There are several instances where the Commander and GNILC simulations are inconsistent with the data fit, with the data result far off in the tails of the distribution with PTEs of 1/300 or 300/300.

This is in contrast with the PTEs for the baseline dust model with $\beta_d$ fixed to 1.55 and $T_d$ fixed to 19.6~K was assumed in the fits and simulations.
With the baseline model, the data lies within the distribution from the simulations for every band and mask region.
Under these circumstances, the \textit{Planck} polarization data can discriminate between those dust models using the residual-dust correlations and prefer a simpler dust model than that produced by the Commander or the GNILC method.

However, this does not necessarily represent a significant difference between these two dust models and the baseline model as we have not incorporated uncertainties on $\beta_d$ or $T_d$ in the analysis, as it is not clear how to analytically propagate them to the PTEs and computationally expensive to numerically estimate. 
Considering the constraints on $\sigma_{\beta_d}$ from Section~\ref{ssec:bdvary}, it is possible that the Commander, GNILC, and baseline dust models are consistent in the masked regions we consider, despite the clear differences in amplitude and structure in Figure~\ref{fig:com_gnilc_bd_maps}.
These results emphasize that the uncertainties on the measured $\beta_d$ are significant and must be accounted for in any analysis of the \textit{Planck} foreground products.

\section{Conclusions}

\label{sec:conc}
A deeper understanding of the polarized dust foreground and its possible spatial variations is motivated by its importance as one of the two dominant polarized foregrounds, along with synchrotron emission, in studies of the polarized CMB.
Although mean values of $\beta_d$ for the high latitude sky have been quoted to high precision,
previous work \citep[e.g.][]{P2018_foregrounds, P2018_dust} suggests that considerable spatial variation in $\beta_d$ cannot be ruled out by the \textit{Planck} data, especially at higher latitudes where the signal to noise ratio of the dust foreground is low.
In this work, we set out to constrain such spatial variations.

We employ two different analyses in map space at a HEALPix resolution of $N_{\mathrm{side}}=16$ ($\sim3.7^\circ$):
a two-band template fitting method and a multifrequency fitting method to analyze the PR3 polarized \textit{Planck} frequency maps and associated data.
We divide our analysis region of about 60\% of the sky into three disjoint regions at high, medium and
low Galactic latitudes (the H, M and L mask regions) each covering about 20\% of the sky (Section~\ref{sec:data}). 
Given the complicated nature of the noise and systematics present in the \textit{Planck} data, our results are calibrated against a suite of 300 simulations based on end-to-end noise simulations provided by \textit{Planck} that include their best estimate of these effects (Section~\ref{ssec:sims}).

The template fit analysis takes the two highest frequency polarization maps at 217 GHz and 353 GHz, which are dominated by the dust foreground, removes \textit{Planck} estimates of the CMB, then performs a dust template fit with the resulting maps to directly estimate $\beta_d$ (Section~\ref{sec:TFmethod}). 
We perform this analysis for our 60\% sky region on the H, M and L regions separately.
By considering these large partitions, we sought to maximize the signal to noise ratio of possible spatial variations in $\beta_d$ at large scales.
Although we test several configurations and partitions for the template fit analysis, all the estimated $\beta_d$ values for all our sky partitions are both consistent within $1\sigma$ with each other and with the \textit{Planck} constraint of a mean $\beta_d = 1.55\pm0.05$ recovered over the full sky \citep{P2018_foregrounds}.

Using our more flexible analysis based on directly fitting a multifrequency foreground model of the CMB, synchrotron foreground, and dust foreground in each pixel, we find the following:
\begin{itemize}
  \item When fitting the CMB and foregrounds together for the polarized \textit{Planck} data, systematics tracing out the \textit{Planck} scan strategy leak into the CMB,  including those CMB maps released by the \textit{Planck} Collaboration. These must be accounted for in any analysis of the \textit{Planck} CMB maps or foreground analyses, including in this paper (Section~\ref{ssec:priors}). These systematics are present even in the latest data releases by the \textit{Planck} Collaboration \citep{P2020}.
  \item Although there are discrepancies between the data and the \textit{Planck} noise simulations in some regions and bands, when taken as a whole, they are consistent over the 60\% of the sky we consider with numerical PTEs of 168 sims/300 sims, 251/300, and 234/300 when compared to our simulation ensemble. The discrepancies in individual regions and bands may be driven by residual systematics in the data that are not completely modeled in the end-to-end simulations provided by \textit{Planck} (Section~\ref{ssec:dvs_comps}).  
  \item Stepping through a central $\beta_d$ assumed when analyzing the data and simulations, we find again that the data prefer $\beta_d \approx 1.55$. The data and our simulations also respond consistently through all central values of $\beta_d$ considered between 1.35 and 1.75, highlighting the consistency of the data and the simulation suite (Section~\ref{ssec:bd_offset}). 
  \item We can set an upper bound for Gaussian variations in $\beta_d$ defined by $\sigma_{\beta_d} = 0.15$ in the lowest Galactic latitude region we consider, $0.30$ at mid-latitudes, and essentially unconstrained at high latitudes, though we find that the data support $\sigma_{\beta_d} \approx{0.45}$.
  These constraints are based on comparisons to 300 simulations from which we estimate numerical PTEs of 297/300, 299/300, 264/300 at these levels of $\sigma_{\beta_d}$ (Section~\ref{ssec:bdvary}). 
  \item We apply the $\sigma_{\beta_d}$ constraints to simulations testing the quality of dust cleaning via template subtraction of \textit{Planck} 353~GHz data that could achieved by future experiments, like \textit{LiteBIRD}. There is an approximately order of magnitude increase in the B-mode residual dust spectrum at $\ell < 32$ due to these allowed levels of $\sigma_{\beta_d}$ versus a spatially constant $\sigma_{\beta_d} = 0$ (Section~\ref{ssec:bdvary}). This should be considered when simulating the performance of future experiments.
  \item Injecting different $\beta_d$ maps derived by the \textit{Planck} Collaboration into the simulations can produce significant differences in our dust variation statistics, shifting numerical PTEs comparing the data against our simulations to 1/300 or 300/300 in many cases where they were in agreement for the baseline case with a constant $\beta_d$. Taking these estimated PTEs at face value, these results suggest that the data reject these alternative spatially varying $\beta_d$ maps in favor of a spatially constant $\beta_d$. The analysis here does not consider errors on the derived $\beta_d$ maps injected into the simulations; so, the dust models tested may be entirely consistent within e.g. the $\sigma_{\beta_d}$ bounds from above. However, these results highlight the need to account for significant uncertainties on the measured $\beta_d$ in any foreground analysis (Section~\ref{ssec:bd_inject}).
\end{itemize}

We have not detected statistically significant spatial variations in $\beta_d$ using the polarized \textit{Planck} data. 
Constraints on $\beta_d$ derived from the two-band template fits (Section~\ref{sec:TFresults}) agree with those published by the \textit{Planck} Collaboration.
A more flexible multifrequency fitting analysis (Section~\ref{sec:fitting_method}) probes the properties of $\beta_d$ in the high latitude regions of the sky and highlights the need for caution when interpreting all-sky mean $\beta_d$ constraints or derived $\beta_d$ maps released by the \textit{Planck} Collaboration.
Even though the mean $\beta_d$ value may be tightly constrained, there is considerable spatial variation allowed by the data.
For example, at high latitudes where the signal to noise ratio of the dust signal is low, the data support $\sigma_{\beta_d}$ at least up to $0.45$.

Variations due to $\sigma_{\beta_d} =$ 0.45 (high-latitude), 0.30 (mid-latitude), or 0.15 (low-latitude) at $N_{\mathrm{side}} = 16$ will impact cosmological parameter recovery and may be important to consider for future experiments targeting the polarized CMB.
Recent simulations for quantifying the performance of future CMB facilities have assumed varying levels of $\sigma_{\beta_d}$ at various scales.
For example, \citet{SO2019} and \citet{CMBS42020} use the \texttt{d1}, \texttt{d4}, and \texttt{d7} dust models from the Python Sky Model (\texttt{PySm}; \citet{PySM2017}), which are all derived from the 2015 \textit{Planck} Commander intensity results \citep{P2015_foregrounds} at $N_{\mathrm{side}} = 128$.
These models assume a lower variation in $\beta_d$ than the polarization data allow.
Given the uncertainty in $\beta_d$ measurements from the current \textit{Planck} polarization data, it may be prudent to consider more conservative dust models when optimizing the design of future surveys like the \texttt{d2} (\texttt{d3}) model in \texttt{PySm} which set $\sigma_{\beta_d}=0.25$ ($0.3$) at degree scales.

\vspace*{0.15in}
We thank Duncan Watts for many helpful comments.
This research was supported in part by NASA grants NNX17AF34G, 80NSSC19K0526, 80NSSC20K0445, and
80NSSC21K0638.
This research has made use of NASA's Astrophysics Data System Bibliographic Services. 
Some of the results in this paper have been derived using the \textsl{healpy} and \textsl{HEALPix} package.
We acknowledge the use of the 
Legacy Archive for Microwave Background Data Analysis (LAMBDA), part of the High Energy Astrophysics Science Archive Center (HEASARC). 
HEASARC/LAMBDA is a service of the Astrophysics Science Division at the NASA Goddard Space Flight Center.  
We also acknowledge use of the \textit{Planck} Legacy Archive. \textit{Planck} is an ESA science mission with instruments and contributions 
directly funded by ESA Member States, NASA, and Canada.

\bibliography{references}

\begin{thebibliography}{}
\expandafter\ifx\csname natexlab\endcsname\relax\def\natexlab#1{#1}\fi
\providecommand{\url}[1]{\href{#1}{#1}}

\bibitem[{{Bennett} {et~al.}(2013){Bennett}, {Larson}, {Weiland}, \&
  {al.}}]{WMAP2013}
{Bennett}, C.~L., {Larson}, D., {Weiland}, J.~L., \& {al.}, e. 2013, \apjs,
  208, 20

\bibitem[{{Bennett} {et~al.}(2003){Bennett}, {Hill}, {Hinshaw}, {Nolta},
  {Odegard}, {Page}, {Spergel}, {Weiland}, {Wright}, {Halpern}, {Jarosik},
  {Kogut}, {Limon}, {Meyer}, {Tucker}, \& {Wollack}}]{wmap2003}
{Bennett}, C.~L., {Hill}, R.~S., {Hinshaw}, G., {et~al.} 2003, \apjs, 148, 97

\bibitem[{{BeyondPlanck Collaboration}(2020)}]{BPlanck_2020}
{BeyondPlanck Collaboration}. 2020, arXiv e-prints, arXiv:2011.05609

\bibitem[{{BICEP2 Collaboration} {et~al.}(2018){BICEP2 Collaboration}, {Keck
  Array Collaboration}, {Ade}, {Ahmed}, {Aikin}, {Alexander}, {Barkats},
  {Benton}, {Bischoff}, {Bock}, {Bowens-Rubin}, {Brevik}, {Buder}, {Bullock},
  {Buza}, {Connors}, {Cornelison}, {Crill}, {Crumrine}, {Dierickx}, {Duband},
  {Dvorkin}, {Filippini}, {Fliescher}, {Grayson}, {Hall}, {Halpern},
  {Harrison}, {Hildebrandt}, {Hilton}, {Hui}, {Irwin}, {Kang}, {Karkare},
  {Karpel}, {Kaufman}, {Keating}, {Kefeli}, {Kernasovskiy}, {Kovac}, {Kuo},
  {Larsen}, {Lau}, {Leitch}, {Lueker}, {Megerian}, {Moncelsi}, {Namikawa},
  {Netterfield}, {Nguyen}, {O'Brient}, {Ogburn}, {Palladino}, {Pryke},
  {Racine}, {Richter}, {Schillaci}, {Schwarz}, {Sheehy}, {Soliman}, {St.
  Germaine}, {Staniszewski}, {Steinbach}, {Sudiwala}, {Teply}, {Thompson},
  {Tolan}, {Tucker}, {Turner}, {Umilt{\`a}}, {Vieregg}, {Wandui}, {Weber},
  {Wiebe}, {Willmert}, {Wong}, {Wu}, {Yang}, {Yoon}, \& {Zhang}}]{BICEP2018}
{BICEP2 Collaboration}, {Keck Array Collaboration}, {Ade}, P.~A.~R., {et~al.}
  2018, \prl, 121, 221301

\bibitem[{{Cardoso} {et~al.}(2008){Cardoso}, {Le Jeune}, {Delabrouille},
  {Betoule}, \& {Patanchon}}]{2008_SMICA}
{Cardoso}, J., {Le Jeune}, M., {Delabrouille}, J., {Betoule}, M., \&
  {Patanchon}, G. 2008, IEEE Journal of Selected Topics in Signal Processing,
  2, 735

\bibitem[{{Chon} {et~al.}(2004){Chon}, {Challinor}, {Prunet}, {Hivon}, \&
  {Szapudi}}]{Polspice2004}
{Chon}, G., {Challinor}, A., {Prunet}, S., {Hivon}, E., \& {Szapudi}, I. 2004,
  \mnras, 350, 914

\bibitem[{{Delouis} {et~al.}(2019){Delouis}, {Pagano}, {Mottet}, {Puget}, \&
  {Vibert}}]{2019_sroll2}
{Delouis}, J.~M., {Pagano}, L., {Mottet}, S., {Puget}, J.~L., \& {Vibert}, L.
  2019, \aap, 629, A38

\bibitem[{{Draine} \& {Hensley}(2013)}]{2013_Draine}
{Draine}, B.~T., \& {Hensley}, B. 2013, \apj, 765, 159

\bibitem[{{Draine} \& {Lee}(1984)}]{Draine1984}
{Draine}, B.~T., \& {Lee}, H.~M. 1984, \apj, 285, 89

\bibitem[{{Dunkley} {et~al.}(2009){Dunkley}, {Spergel}, {Komatsu}, {Hinshaw},
  {Larson}, {Nolta}, {Odegard}, {Page}, {Bennett}, {Gold}, {Hill}, {Jarosik},
  {Weiland}, {Halpern}, {Kogut}, {Limon}, {Meyer}, {Tucker}, {Wollack}, \&
  {Wright}}]{2009_Dunkley}
{Dunkley}, J., {Spergel}, D.~N., {Komatsu}, E., {et~al.} 2009, \apj, 701, 1804

\bibitem[{Dvoretzky {et~al.}(1956)Dvoretzky, Kiefer, \& Wolfowitz}]{DKW}
Dvoretzky, A., Kiefer, J., \& Wolfowitz, J. 1956, Ann. Math. Statist., 27, 642.
\newblock \url{https://doi.org/10.1214/aoms/1177728174}

\bibitem[{{Eriksen} {et~al.}(2004){Eriksen}, {O'Dwyer}, {Jewell}, {Wand elt},
  {Larson}, {G{\'o}rski}, {Levin}, {Banday}, \& {Lilje}}]{2004_COM}
{Eriksen}, H.~K., {O'Dwyer}, I.~J., {Jewell}, J.~B., {et~al.} 2004, \apjs, 155,
  227

\bibitem[{{Fern{\'a}ndez-Cobos} {et~al.}(2012){Fern{\'a}ndez-Cobos}, {Vielva},
  {Barreiro}, \& {Mart{\'\i}nez-Gonz{\'a}lez}}]{2012_SEVEM}
{Fern{\'a}ndez-Cobos}, R., {Vielva}, P., {Barreiro}, R.~B., \&
  {Mart{\'\i}nez-Gonz{\'a}lez}, E. 2012, \mnras, 420, 2162

\bibitem[{{Fuskeland} {et~al.}(2021){Fuskeland}, {Andersen}, {Aurlien},
  {Banerji}, {Brilenkov}, {Eriksen}, {Galloway}, {Gjerl{\o}w}, {N{\ae}ss},
  {Svalheim}, \& {Wehus}}]{Fuskeland2021}
{Fuskeland}, U., {Andersen}, K.~J., {Aurlien}, R., {et~al.} 2021, \aap, 646,
  A69

\bibitem[{{G{\'o}rski} {et~al.}(2005){G{\'o}rski}, {Hivon}, {Banday}, {Wand
  elt}, {Hansen}, {Reinecke}, \& {Bartelmann}}]{2005_Healpix}
{G{\'o}rski}, K.~M., {Hivon}, E., {Banday}, A.~J., {et~al.} 2005, \apj, 622,
  759

\bibitem[{{Hazumi} {et~al.}(2020){Hazumi}, {Ade}, {Adler}, {Allys}, {Arnold},
  {Auguste}, {Aumont}, {Aurlien}, {Austermann}, {Baccigalupi}, {Banday},
  {Banjeri}, {Barreiro}, {Basak}, {Beall}, {Beck}, {Beckman}, {Bermejo}, {de
  Bernardis}, {Bersanelli}, {Bonis}, {Borrill}, {Boulanger}, {Bounissou},
  {Brilenkov}, {Brown}, {Bucher}, {Calabrese}, {Campeti}, {Carones}, {Casas},
  {Challinor}, {Chan}, {Cheung}, {Chinone}, {Cliche}, {Colombo}, {Columbro},
  {Cubas}, {Cukierman}, {Curtis}, {D'Alessandro}, {Dachlythra}, {De Petris},
  {Dickinson}, {Diego-Palazuelos}, {Dobbs}, {Dotani}, {Duband}, {Duff},
  {Duval}, {Ebisawa}, {Elleflot}, {Eriksen}, {Errard}, {Essinger-Hileman},
  {Finelli}, {Flauger}, {Franceschet}, {Fuskeland}, {Galloway}, {Ganga}, {Gao},
  {Genova-Santos}, {Gerbino}, {Gervasi}, {Ghigna}, {Gjerl{\o}w}, {Gradziel},
  {Grain}, {Grupp}, {Gruppuso}, {Gudmundsson}, {de Haan}, {Halverson},
  {Hargrave}, {Hasebe}, {Hasegawa}, {Hattori}, {Henrot-Versill{\'e}}, {Herman},
  {Herranz}, {Hill}, {Hilton}, {Hirota}, {Hivon}, {Hlozek}, {Hoshino}, {de la
  Hoz}, {Hubmayr}, {Ichiki}, {Iida}, {Imada}, {Ishimura}, {Ishino}, {Jaehnig},
  {Kaga}, {Kashima}, {Katayama}, {Kato}, {Kawasaki}, {Keskitalo}, {Kisner},
  {Kobayashi}, {Kogiso}, {Kogut}, {Kohri}, {Komatsu}, {Komatsu}, {Konishi},
  {Krachmalnicoff}, {Kreykenbohm}, {Kuo}, {Kushino}, {Lamagna}, {Lanen},
  {Lattanzi}, {Lee}, {Leloup}, {Levrier}, {Linder}, {Louis}, {Luzzi},
  {Maciaszek}, {Maffei}, {Maino}, {Maki}, {Mandelli}, {Martinez-Gonzalez},
  {Masi}, {Matsumura}, {Mennella}, {Migliaccio}, {Minami}, {Mitsuda},
  {Montgomery}, {Montier}, {Morgante}, {Mot}, {Murata}, {Murphy}, {Nagai},
  {Nagano}, {Nagasaki}, {Nagata}, {Nakamura}, {Namikawa}, {Natoli}, {Nerval},
  {Nishibori}, {Nishino}, {Noviello}, {O'Sullivan}, {Ogawa}, {Ogawa}, {Oguri},
  {Ohsaki}, {Ohta}, {Okada}, {Okada}, {Pagano}, {Paiella}, {Paoletti},
  {Patanchon}, {Peloton}, {Piacentini}, {Pisano}, {Polenta}, {Poletti},
  {Prouv{\'e}}, {Puglisi}, {Rambaud}, {Raum}, {Realini}, {Reinecke},
  {Remazeilles}, {Ritacco}, {Roudil}, {Rubino-Martin}, {Russell}, {Sakurai},
  {Sakurai}, {Sandri}, {Sasaki}, {Savini}, {Scott}, {Seibert}, {Sekimoto},
  {Sherwin}, {Shinozaki}, {Shiraishi}, {Shirron}, {Signorelli}, {Smecher},
  {Stever}, {Stompor}, {Sugai}, {Sugiyama}, {Suzuki}, {Suzuki}, {Svalheim},
  {Switzer}, {Takaku}, {Takakura}, {Takakura}, {Takase}, {Takeda}, {Tartari},
  {Taylor}, {Terao}, {Thommesen}, {Thompson}, {Thorne}, {Toda}, {Tomasi},
  {Tominaga}, {Trappe}, {Tristram}, {Tsuji}, {Tsujimoto}, {Tucker}, {Ullom},
  {Vermeulen}, {Vielva}, {Villa}, {Vissers}, {Vittorio}, {Wehus}, {Weller},
  {Westbrook}, {Wilms}, {Winter}, {Wollack}, {Yamasaki}, {Yoshida}, {Yumoto},
  {Zannoni}, \& {Zonca}}]{Hazumi2020}
{Hazumi}, M., {Ade}, P.~A.~R., {Adler}, A., {et~al.} 2020, in Society of
  Photo-Optical Instrumentation Engineers (SPIE) Conference Series, Vol. 11443,
  Society of Photo-Optical Instrumentation Engineers (SPIE) Conference Series,
  114432F

\bibitem[{{Hildebrand} {et~al.}(2000){Hildebrand}, {Davidson}, {Dotson},
  {Dowell}, {Novak}, \& {Vaillancourt}}]{Hildebrand2000}
{Hildebrand}, R.~H., {Davidson}, J.~A., {Dotson}, J.~L., {et~al.} 2000, \pasp,
  112, 1215

\bibitem[{{Kamionkowski} {et~al.}(1997){Kamionkowski}, {Kosowsky}, \&
  {Stebbins}}]{kamionkowski1997}
{Kamionkowski}, M., {Kosowsky}, A., \& {Stebbins}, A. 1997, \prd, 55, 7368

\bibitem[{{Krachmalnicoff} {et~al.}(2018){Krachmalnicoff}, {Carretti},
  {Baccigalupi}, {Bernardi}, {Brown}, {Gaensler}, {Haverkorn}, {Kesteven},
  {Perrotta}, {Poppi}, \& {Staveley-Smith}}]{Krachmalnicoff2018}
{Krachmalnicoff}, N., {Carretti}, E., {Baccigalupi}, C., {et~al.} 2018, \aap,
  618, A166

\bibitem[{{Lazarian}(2007)}]{Lazarian2009}
{Lazarian}, A. 2007, \jqsrt, 106, 225

\bibitem[{{Lazarian} \& {Finkbeiner}(2003)}]{Lazarian2003}
{Lazarian}, A., \& {Finkbeiner}, D. 2003, \nar, 47, 1107

\bibitem[{{Leach} {et~al.}(2008){Leach}, {Cardoso}, {Baccigalupi}, {Barreiro},
  {Betoule}, {Bobin}, {Bonaldi}, {Delabrouille}, {de Zotti}, {Dickinson},
  {Eriksen}, {Gonz{\'a}lez-Nuevo}, {Hansen}, {Herranz}, {Le Jeune},
  {L{\'o}pez-Caniego}, {Mart{\'\i}nez-Gonz{\'a}lez}, {Massardi}, {Melin},
  {Miville-Desch{\^e}nes}, {Patanchon}, {Prunet}, {Ricciardi}, {Salerno},
  {Sanz}, {Starck}, {Stivoli}, {Stolyarov}, {Stompor}, \& {Vielva}}]{sevem2008}
{Leach}, S.~M., {Cardoso}, J.~F., {Baccigalupi}, C., {et~al.} 2008, \aap, 491,
  597

\bibitem[{{Mart{\'\i}nez-Gonz{\'a}lez}
  {et~al.}(2003){Mart{\'\i}nez-Gonz{\'a}lez}, {Diego}, {Vielva}, \&
  {Silk}}]{sevem2003}
{Mart{\'\i}nez-Gonz{\'a}lez}, E., {Diego}, J.~M., {Vielva}, P., \& {Silk}, J.
  2003, \mnras, 345, 1101

\bibitem[{{Pagano} {et~al.}(2020){Pagano}, {Delouis}, {Mottet}, {Puget}, \&
  {Vibert}}]{Pagano2020}
{Pagano}, L., {Delouis}, J.~M., {Mottet}, S., {Puget}, J.~L., \& {Vibert}, L.
  2020, \aap, 635, A99

\bibitem[{{Page} {et~al.}(2007){Page}, {Hinshaw}, {Komatsu}, {Nolta},
  {Spergel}, {Bennett}, {Barnes}, {Bean}, {Dor{\'e}}, {Dunkley}, {Halpern},
  {Hill}, {Jarosik}, {Kogut}, {Limon}, {Meyer}, {Odegard}, {Peiris}, {Tucker},
  {Verde}, {Weiland}, {Wollack}, \& {Wright}}]{Page2007}
{Page}, L., {Hinshaw}, G., {Komatsu}, E., {et~al.} 2007, \apjs, 170, 335

\bibitem[{{Pelgrims} {et~al.}(2021){Pelgrims}, {Clark}, {Hensley},
  {Panopoulou}, {Pavlidou}, {Tassis}, {Eriksen}, \& {Wehus}}]{Pelgrims2021}
{Pelgrims}, V., {Clark}, S.~E., {Hensley}, B.~S., {et~al.} 2021, \aap, 647, A16

\bibitem[{{Penzias} \& {Wilson}(1965)}]{PW1965}
{Penzias}, A.~A., \& {Wilson}, R.~W. 1965, \apj, 142, 419

\bibitem[{{\sorthelp{Planck Collaboration 2014I}}{Planck Collaboration
  IX}(2014)}]{P2013HFIcalib}
{\sorthelp{Planck Collaboration 2014I}}{Planck Collaboration IX}. 2014, \aap,
  571, A9

\bibitem[{{\sorthelp{Planck Collaboration 2015A}}{Planck Collaboration
  I}(2016)}]{P2016}
{\sorthelp{Planck Collaboration 2015A}}{Planck Collaboration I}. 2016, \aap,
  594, A1

\bibitem[{{\sorthelp{Planck Collaboration 2015J}}{Planck Collaboration
  X}(2016)}]{P2015_foregrounds}
{\sorthelp{Planck Collaboration 2015J}}{Planck Collaboration X}. 2016, \aap,
  594, A10

\bibitem[{{\sorthelp{Planck Collaboration 2015M}}{Planck Collaboration
  XIII}(2016)}]{P2015_results}
{\sorthelp{Planck Collaboration 2015M}}{Planck Collaboration XIII}. 2016, \aap,
  594, A13

\bibitem[{{\sorthelp{Planck Collaboration 2018B}}{Planck Collaboration
  II}(2019)}]{P2018_LFI}
{\sorthelp{Planck Collaboration 2018B}}{Planck Collaboration II}. 2019, \aap,
  in press, arXiv:1807.06206

\bibitem[{{\sorthelp{Planck Collaboration 2018C}}{Planck Collaboration
  III}(2019)}]{P2018_HFI}
{\sorthelp{Planck Collaboration 2018C}}{Planck Collaboration III}. 2019, \aap,
  in press, arXiv:1807.06207

\bibitem[{{\sorthelp{Planck Collaboration 2018D}}{Planck Collaboration
  IV}(2019)}]{P2018_foregrounds}
{\sorthelp{Planck Collaboration 2018D}}{Planck Collaboration IV}. 2019, \aap,
  in press, arXiv:1807.06208

\bibitem[{{\sorthelp{Planck Collaboration 2018F}}{Planck Collaboration
  VI}(2019)}]{P2018_params}
{\sorthelp{Planck Collaboration 2018F}}{Planck Collaboration VI}. 2019, \aap,
  in press, arXiv:1807.06209

\bibitem[{{\sorthelp{Planck Collaboration 2018K}}{Planck Collaboration
  XI}(2019)}]{P2018_dust}
{\sorthelp{Planck Collaboration 2018K}}{Planck Collaboration XI}. 2019, \aap,
  in press, arXiv:1801.04945

\bibitem[{{\sorthelp{Planck Collaboration IntV}}{Planck Collaboration Int.
  XXII}(2015)}]{P2015_dust}
{\sorthelp{Planck Collaboration IntV}}{Planck Collaboration Int. XXII}. 2015,
  \aap, 576, A107

\bibitem[{{\sorthelp{Planck Collaboration IntZY}}{Planck Collaboration Int.
  L}(2017)}]{P2016_dust}
{\sorthelp{Planck Collaboration IntZY}}{Planck Collaboration Int. L}. 2017,
  \aap, 599, A51

\bibitem[{{\sorthelp{Planck Collaboration IntZZG}}{Planck Collaboration Int.
  LVII}(2020)}]{P2020}
{\sorthelp{Planck Collaboration IntZZG}}{Planck Collaboration Int. LVII}. 2020,
  \aap, submitted

\bibitem[{{Remazeilles} {et~al.}(2011){Remazeilles}, {Delabrouille}, \&
  {Cardoso}}]{2011_GNILC}
{Remazeilles}, M., {Delabrouille}, J., \& {Cardoso}, J.-F. 2011, \mnras, 418,
  467

\bibitem[{{Seljebotn} {et~al.}(2017){Seljebotn}, {B{\ae}rland}, {Eriksen},
  {Mardal}, \& {Wehus}}]{2017_COM}
{Seljebotn}, D.~S., {B{\ae}rland}, T., {Eriksen}, H.~K., {Mardal}, K.~A., \&
  {Wehus}, I.~K. 2017, arXiv e-prints, arXiv:1710.00621

\bibitem[{{Sheehy} \& {Slosar}(2018)}]{Sheehy2018}
{Sheehy}, C., \& {Slosar}, A. 2018, \prd, 97, 043522

\bibitem[{{Simons Observatory Collaboration}(2019)}]{SO2019}
{Simons Observatory Collaboration}. 2019, \jcap, 2019, 056

\bibitem[{{The CMB-S4 Collaboration}(2020)}]{CMBS42020}
{The CMB-S4 Collaboration}. 2020, arXiv e-prints, arXiv:2008.12619

\bibitem[{{Thorne} {et~al.}(2017){Thorne}, {Dunkley}, {Alonso}, \&
  {N{\ae}ss}}]{PySM2017}
{Thorne}, B., {Dunkley}, J., {Alonso}, D., \& {N{\ae}ss}, S. 2017, \mnras, 469,
  2821

\bibitem[{{Tristram} {et~al.}(2021){Tristram}, {Banday}, {G{\'o}rski},
  {Keskitalo}, {Lawrence}, {Andersen}, {Barreiro}, {Borrill}, {Eriksen},
  {Fernandez-Cobos}, {Kisner}, {Mart{\'\i}nez-Gonz{\'a}lez}, {Partridge},
  {Scott}, {Svalheim}, {Thommesen}, \& {Wehus}}]{Tristam2021}
{Tristram}, M., {Banday}, A.~J., {G{\'o}rski}, K.~M., {et~al.} 2021, \aap, 647,
  A128

\bibitem[{{Watts} {et~al.}(2020){Watts}, {Addison}, {Bennett}, \&
  {Weiland}}]{Watts2020}
{Watts}, D.~J., {Addison}, G.~E., {Bennett}, C.~L., \& {Weiland}, J.~L. 2020,
  \apj, 889, 130

\bibitem[{{Weiland} {et~al.}(2018){Weiland}, {Osumi}, {Addison}, {Bennett},
  {Watts}, {Halpern}, \& {Hinshaw}}]{Weiland2018}
{Weiland}, J.~L., {Osumi}, K., {Addison}, G.~E., {et~al.} 2018, \apj, 863, 161

\bibitem[{{Weingartner} \& {Draine}(2001)}]{Weingartner2001}
{Weingartner}, J.~C., \& {Draine}, B.~T. 2001, \apj, 548, 296

\end{thebibliography}

\end{document}